\documentclass[12pt]{iopart}
\usepackage{graphicx}

\usepackage{iopams}
\usepackage{braket}

\usepackage{tikz}
\usetikzlibrary{cd}
\usetikzlibrary{calc}
\tikzset{anchorbase/.style={baseline={([yshift=-0.5ex]current bounding box.center)}},
  tinynodes/.style={font=\tiny, text height=0.25ex, text depth=0.05ex},
  smallnodes/.style={font=\scriptsize, text height=0.75ex, text depth=0.15ex},
  busual/.style={line width=1.0,color=blueberry},
  usual/.style={line width=0.9,color=black},
  rusual/.style={line width=1.0,color=dark-red,double},
  crossline/.style={preaction={draw=white,line width=5.0pt,-}},
  scrossline/.style={preaction={draw=white,line width=4.0pt,-}},
}

\usetikzlibrary{decorations.pathreplacing}

\usepackage{ytableau}

\definecolor{dark-red}{rgb}{0.7,0.25,0.25}

\definecolor{boxfill}{rgb}{0.64, 0.64, 0.82} 

\newcommand{\genh}{3}

\newcommand{\scale}{.7} 

\newcommand{\scaleb}{.8}

\usetikzlibrary{matrix, arrows, calc}
\newcommand{\ru}{to [out=0,in=270]}
\newcommand{\rd}{to [out=0,in=90]}
\newcommand{\ur}{to [out=90,in=180]}
\newcommand{\ul}{to [out=90,in=0]}
\newcommand{\lu}{to [out=180,in=270]}

\newcommand{\dr}{to [out=270,in=180]}

 \newcommand{\pgen}[3]{
\draw[usual] (#1, 0) to (#1+ 0.5, \genh/6); 
\draw[usual] (#1+1, 0) to (#1+ 0.5, \genh/6); 
\draw[usual] (#1, \genh/2) to (#1+ 0.5, 2*\genh/6); 
\draw[usual] (#1+1, \genh/2) to (#1+ 0.5, 2*\genh/6); 
\draw[usual] (#1+0.5, \genh/6) to (#1+ 0.5, 2*\genh/6);
\node at (#1,-.5) {\tiny #2};
\node at (#1+1,-.5) {\tiny #3};
}

 \newcommand{\sgen}[3]{
	\draw[usual] (#1, 0) to (#1+ \genh/6, \genh/4); 
	\draw[usual] (#1+1.5, 0) to (#1+1.5- \genh/6, \genh/4); 
	\draw[usual] (#1, \genh/2) to (#1+ \genh/6, \genh/4); 
	\draw[usual] (#1+1.5, \genh/2) to (#1+1.5- \genh/6, \genh/4); 
	\draw[usual] (#1+\genh/6, \genh/4) to (#1+1.5- \genh/6, \genh/4);
	\node at (#1,-.5) {\tiny #2};
	\node at (#1+1.5,-.5) {\tiny #3};
}

\newcommand{\egen}[3]{
	\draw[usual] (#1,0) \ur (#1+.5,\genh/6) \rd (#1+1,0);
	\draw[usual] (#1,\genh/2) \dr (#1+.5,\genh/2-\genh/6) \ru (#1+1,\genh/2);
	\node at (#1,-.5) {\scalebox{\scale}{#2}};
	\node at (#1+1,-.5) {\scalebox{\scale}{#3}};
}

\newcommand{\idgen}[3]{
	\draw[usual] (#1,0) \ru (#1+\genh/6,\genh/4) \ul (#1,\genh/2);
	\draw[usual] (#1 + 1.5,0) \lu (#1+1.5 - \genh/6,\genh/4) \ur (#1+1.5,\genh/2);
	\node at (#1,-.5) {\scalebox{\scale}{#2}};
	\node at (#1+1.5,-.5) {\scalebox{\scale}{#3}};
}

\newcommand{\twobox}[2]{
	\filldraw[usual, draw=black, fill = boxfill] (#1-0.25, \genh/3 -0.5 -0.25) rectangle (#1 +1.25, \genh/3-0.25)node [pos=.5]{\scalebox{\scale}{#2}};
	\filldraw[black] (#1-0.25, \genh/3-0.25-0.25) circle (1pt);
}

\newcommand{\twoboxs}[2]{
	\filldraw[usual, draw=black, fill = boxfill] (#1-0.25, 3*\genh/8-0.25) rectangle (#1 +1.25, 3*\genh/8+0.25)node [pos=.5]{\scalebox{\scale}{#2}};
	\filldraw[black] (#1-0.25, 3*\genh/8) circle (1pt);
}

\newcommand{\twoboxr}[2]{
	\draw[usual] (#1, 0) to [out = 90, in = 180] (#1+0.35, 1);
	\draw[usual] (#1+ 0.35, 2) to [out = -180, in =-90] (#1, \genh);
	\draw[usual] (#1+1, 0) to [out = 90, in = 0] (#1+ 0.65, 1);
	\draw[usual] (#1+.65, 2) to [out = 0, in =-90] (#1+1, \genh);
	\filldraw[usual, draw=black, fill = boxfill] (#1+0.25, 0.75) rectangle (#1 +0.75, 2.25)node [pos=.5]{\scalebox{\scale}{#2}};
	\filldraw[black] (#1+0.5, 0.75) circle (1pt);
}

\newcommand{\twoboxrs}[2]{
	\draw[usual] (#1, 0) to [out = 90, in = 180] (#1+0.35, \genh/4);
	\draw[usual] (#1+ 0.35, \genh/2) to [out = -180, in =-90] (#1, 3*\genh/4);
	\draw[usual] (#1+1, 0) to [out = 90, in = 0] (#1+1- 0.35, \genh/4);
	\draw[usual] (#1+.65, \genh/2) to [out = 0, in =-90] (#1+1, 3*\genh/4);
	\filldraw[usual, draw=black, fill = boxfill] (#1+0.25, 3*\genh/8 -0.75) rectangle (#1 +0.75, 3*\genh/8 +0.75 )node [pos=.5]{\scalebox{\scale}{#2}};
	\filldraw[black] (#1+0.5, 3*\genh/8 -0.75) circle (1pt);
}

\newcommand{\twoboxb}[2]{
	\filldraw[usual, draw=black, fill = boxfill] (#1-0.25, \genh/3 -0.5) rectangle (#1 +1.25, \genh/3)node [pos=.5]{\scalebox{\scale}{#2}};
	\filldraw[black] (#1-0.25, \genh/3-0.25) circle (1pt);
}

\newcommand{\shortcup}[3]{
	\draw[usual] (#1,\genh/3) \dr (#1+.5,\genh/6) \ru (#1+1,\genh/3);
	\node at (#1,0.5) {\tiny #2};
	\node at (#1+1,.5) {\tiny #3};
}

\newcommand{\shortcap}[3]{
	\draw[usual] (#1,0) \ur (#1+.5,\genh/6) \rd (#1+1,0);
	\node at (#1,-.5) {\tiny #2};
	\node at (#1+1,-.5) {\tiny #3};
}

\newcommand{\su}[2]{
  \draw[usual,crossline] (#1,0) to (#1,\genh);
  \node at (#1,-.5) {\scalebox{\scale}{#2}};
}

\newcommand{\suhb}[2]{
	\draw[usual,crossline] (#1,0) to (#1,3*\genh/4);
	\node at (#1,-.5) {\scalebox{\scale}{#2}};
}

\newcommand{\suhd}[2]{
	\draw[usual,crossline] (#1,0) to (#1,2*\genh/3);
	\node at (#1,-.5) {\scalebox{\scale}{#2}};
}

\newcommand{\sutl}[2]{
	\draw[usual,crossline] (#1,0) to (#1,5*\genh/6);
	\node at (#1,-.5) {\scalebox{\scale}{#2}};
}

\newcommand{\cupcap}[3]{
  \draw[usual] (#1,0) \ur (#1+.5,\genh/3) \rd (#1+1,0);
  \draw[usual] (#1,\genh) \dr (#1+.5,\genh-\genh/3) \ru (#1+1,\genh);
  \node at (#1,-.5) {\scalebox{\scale}{#2}};
  \node at (#1+1,-.5) {\scalebox{\scale}{#3}};
}

\newcommand{\cupcaphb}[3]{
	\draw[usual] (#1,0) \ur (#1+.5,\genh/4) \rd (#1+1,0);
	\draw[usual] (#1,3*\genh/4 ) \dr (#1+.5,2*\genh/4) \ru (#1+1,3*\genh/4);
	\node at (#1,-.5) {\scalebox{\scale}{#2}};
	\node at (#1+1,-.5) {\scalebox{\scale}{#3}};
} 

\newcommand{\cupcapb}[3]{
	\draw[usual] (#1,0) \ur (#1+.5,\genh/9) \rd (#1+1,0);
	\draw[usual] (#1,\genh/3) \dr (#1+.5,\genh/3-\genh/9) \ru (#1+1,\genh/3);
	\node at (#1,-.5) {\scalebox{\scale}{#2}};
	\node at (#1+1,-.5) {\scalebox{\scale}{#3}};
}

\newcommand{\kbox}[3]{
	\su{#1}{~};
	\filldraw[usual, fill = boxfill] (#1-\genh/24, \genh/2 -\genh/24) rectangle (#1+\genh/24, \genh/2 + \genh/24);
	\node at (#1,-.5) {\scalebox{\scale}{ #2}};
	\node[anchor = east] at (#1, \genh/2) {\scalebox{\scale}{ #3}};
}

\newcommand{\kboxw}[3]{
	\su{#1}{~};
	\filldraw[usual, fill = boxfill] (#1-\genh/24, \genh/2 -\genh/24) rectangle (#1+\genh/24, \genh/2 + \genh/24);
	\node at (#1,-.5) {\scalebox{\scaleb}{ #2}};
	\node[anchor=east] at (#1, \genh/2) {\scalebox{\scale}{ #3}};
}

\newcommand{\kboxh}[3]{
	\sub{#1}{~};
	\filldraw[usual, fill = boxfill] (#1-\genh/24, \genh/6 -\genh/24) rectangle (#1+\genh/24, \genh/6 + \genh/24);
	\node[anchor = east] at (#1, \genh/6) {\scalebox{\scale}{#3}};
	\node at (#1,-.5) {\scalebox{\scale}{#2}};
} 

\newcommand{\kboxhd}[3]{
	\draw[usual] (#1,0) -- (#1, 2*\genh/3);
	\filldraw[usual, fill = boxfill] (#1-\genh/24, \genh/3 -\genh/24) rectangle (#1+\genh/24, \genh/3 + \genh/24);
	\node[anchor = east]  at (#1, \genh/3) {\scalebox{\scale}{#3}};
	\node at (#1,-.5) {\scalebox{\scale}{#2}};
} 

\newcommand{\kboxt}[3]{
	\sub{#1}{~};
	\filldraw[usual, fill =boxfill] (#1-\genh/24, 2*\genh/9 -\genh/24) rectangle (#1+\genh/24,2*\genh/9 + \genh/24);
	\node[anchor = east]  at (#1, 2*\genh/9) {\scalebox{\scale}{#3}};
	\node at (#1,-.5) {\scalebox{\scale}{#2}};
}

\newcommand{\kboxb}[3]{
	\sub{#1}{~};
	\filldraw[usual, fill = boxfill] (#1-\genh/24, \genh/9 -\genh/24) rectangle (#1+\genh/24, \genh/9 + \genh/24);
	\node[anchor = east] at (#1, \genh/9) {\scalebox{\scale}{#3}};
	\node at (#1,-.5) {\scalebox{\scale}{#2}};
}

\newcommand{\kboxtl}[3]{
	\su{#1}{~};
	\filldraw[usual, fill = boxfill] (#1-\genh/24, 2*\genh/3 -\genh/24) rectangle (#1+\genh/24,2*\genh/3 + \genh/24);
	\node[anchor = east]  at (#1, 2*\genh/3) {\scalebox{\scale}{#3}};
	\node at (#1,-.5) {\scalebox{\scale}{#2}};
}

\newcommand{\kboxbl}[3]{
	\su{#1}{~};
	\filldraw[usual, fill = boxfill] (#1-\genh/24, \genh/3 -\genh/24) rectangle (#1+\genh/24, \genh/3 + \genh/24);
	\node[anchor = east]  at (#1, \genh/3) {\scalebox{\scale}{#3}};
	\node at (#1,-.5) {\scalebox{\scale}{#2}};
}

\newcommand{\kboxd}[4]{
	\su{#1}{~};
	\filldraw[usual, fill = boxfill] (#1-\genh/24, \genh/3 -\genh/24) rectangle (#1+\genh/24, \genh/3 + \genh/24);
	\filldraw[usual, fill = boxfill] (#1-\genh/24, 2*\genh/3 -\genh/24) rectangle (#1+\genh/24,2*\genh/3 + \genh/24);
	\node at (#1,-.5) {\scalebox{\scale}{#2}};
	\node[anchor = east]  at (#1, 2*\genh/3) {\scalebox{\scale}{#3}}; 
	\node[anchor = east]  at (#1, \genh/3) {\scalebox{\scale}{#4}}; 
}

\newcommand{\boxcupcapbox}[5]{
	\draw[usual] (#1,\genh) \dr (#1+0.5,\genh-\genh/3) \ru (#1+1,\genh);
	\draw[usual] (#1,0) \ur (#1+.5,\genh/3) \rd (#1+1,0);
	\filldraw[usual, draw=black, fill = boxfill] (#1-\genh/24, \genh - \genh/6 - \genh/24) rectangle (#1+\genh/24,\genh -  \genh/6 + \genh/24);
	\filldraw[usual, draw=black, fill = boxfill] (#1-\genh/24, \genh/6 - \genh/24) rectangle (#1+\genh/24, \genh/6 + \genh/24);
	\node at (#1,-.5) {\scalebox{\scale}{#2}};
	\node at (#1+1,-.5) {\scalebox{\scale}{ #3}};
	\node[anchor = east]  at (#1, \genh - \genh/6) {\scalebox{\scale}{#4}};
	\node[anchor = east]  at (#1, \genh/6) {\scalebox{\scale}{#5}};
}

\newcommand{\boxcupcapboxhb}[5]{
	\draw[usual] (#1,0) \ur (#1+.5,\genh/4) \rd (#1+1,0);
	\draw[usual] (#1,3*\genh/4 ) \dr (#1+.5,2*\genh/4) \ru (#1+1,3*\genh/4);
	\filldraw[usual, draw=black, fill = boxfill] (#1-\genh/24, 3*\genh/4 - \genh/8 - \genh/24) rectangle (#1+\genh/24,3*\genh/4 - \genh/8+ \genh/24);
	\filldraw[usual, draw=black, fill = boxfill] (#1-\genh/24, \genh/8- \genh/24) rectangle (#1+\genh/24, \genh/8 + \genh/24);
	\node at (#1,-.5) {\scalebox{\scale}{#2}};
	\node at (#1+1,-.5) {\scalebox{\scale}{ #3}};
	\node[anchor = east] at (#1, 3*\genh/4 - \genh/8) {\scalebox{\scale}{#4}};
	\node[anchor = east] at (#1, \genh/8) {\scalebox{\scale}{#5}};
} 

\newcommand{\smallboxcupcapbox}[5]{
	\draw[usual] (#1,\genh- \genh/6) \dr (#1+0.5,\genh-\genh/3- \genh/6) \ru (#1+1,\genh- \genh/6);
	\draw[usual] (#1,0) \ur (#1+.5,\genh/3) \rd (#1+1,0);
	\filldraw[usual, draw=black, fill = boxfill] (#1-\genh/24, \genh- \genh/6 - \genh/6 - \genh/24 ) rectangle (#1+\genh/24,\genh- \genh/6 -  \genh/6 + \genh/24);
	\filldraw[usual, draw=black, fill = boxfill] (#1-\genh/24, \genh/6 - \genh/24) rectangle (#1+\genh/24, \genh/6 + \genh/24);
	\node at (#1,-.3) {\scalebox{\scale}{#2}};
	\node at (#1+1,-.3) {\scalebox{\scale}{ #3}};
	\node[anchor = east]  at (#1, \genh - \genh/6- \genh/6) {\scalebox{\scale}{#4}};
	\node[anchor = east] at (#1, \genh/6) {\scalebox{\scale}{#5}};
}

\newcommand{\hboxcupcapbox}[5]{
	\draw[usual] (#1,\genh) \dr (#1+0.5,\genh-\genh/3) \ru (#1+1,\genh);
	\draw[usual] (#1,0) \ur (#1+.5,\genh/3) \rd (#1+1,0);
	\filldraw[usual, draw=black, fill = boxfill] (#1-\genh/24, \genh - \genh/6 - \genh/24) rectangle (#1+\genh/24,\genh -  \genh/6 + \genh/24);
	\filldraw[usual, draw=black, fill = boxfill] (#1-\genh/24, \genh/6 - \genh/24) rectangle (#1+\genh/24, \genh/6 + \genh/24);
	\node at (#1,-.5) {\scalebox{\scale}{#2}};
	\node at (#1+1,-.5) {\scalebox{\scale}{#3}};
	\node[anchor = east] at (#1, \genh - \genh/6) {\scalebox{\scale}{#4}};
	\node[anchor = east] at (#1, \genh/6) {\scalebox{\scale}{#5}};
}

\newcommand{\eboxcupcapbox}[5]{
	\draw[usual] (#1,\genh) \dr (#1+0.5,\genh-\genh/3) \ru (#1+1,\genh);
	\draw[usual] (#1,0) \ur (#1+.5,\genh/3) \rd (#1+1,0);
	\filldraw[usual, draw=black, fill =boxfill] (#1-\genh/24, \genh - \genh/6 - \genh/24) rectangle (#1+\genh/24,\genh -  \genh/6 + \genh/24);
	\filldraw[usual, draw=black, fill = boxfill] (#1-\genh/24, \genh/6 - \genh/24) rectangle (#1+\genh/24, \genh/6 + \genh/24);
	\node at (#1,-.5) {\scalebox{\scale}{#2}};
	\node at (#1+1,-.5) {\scalebox{\scale}{#3}};
	\node[anchor = east] at (#1, \genh - \genh/6) {\scalebox{\scale}{#4}};
	\node[anchor = east] at (#1, \genh/6) {\scalebox{\scale}{#5}};
}

\newcommand{\ebboxcupcapbox}[5]{
	\draw[usual] (#1,\genh) \dr (#1+0.5,\genh-\genh/3) \ru (#1+1,\genh);
	\draw[usual] (#1,0) \ur (#1+.5,\genh/3) \rd (#1+1,0);
	\filldraw[usual, draw=black, fill = boxfill] (#1+1-\genh/24, \genh - \genh/6 - \genh/24) rectangle (#1+1+\genh/24,\genh -  \genh/6 + \genh/24);
	\filldraw[usual, draw=black, fill = boxfill] (#1+1-\genh/24, \genh/6 - \genh/24) rectangle (#1+1+\genh/24, \genh/6 + \genh/24);
	\node at (#1,-.5) {\scalebox{\scale}{#2}};
	\node at (#1+1,-.5) {\scalebox{\scale}{#3}};
	\node[anchor = east] at (#1+1, \genh - \genh/6) {\scalebox{\scale}{#4}};
	\node[anchor = east] at (#1+1, \genh/6) {\scalebox{\scale}{#5}};
} 

\newcommand{\boxcup}[4]{
	\draw[usual] (#1,\genh/3) \dr (#1+0.5,\genh/3-\genh/3) \ru (#1+1,\genh/3);
	\filldraw[usual, draw=black, fill = boxfill] (#1-\genh/24, \genh/3 - \genh/6 - \genh/24) rectangle (#1+\genh/24,\genh/3 -  \genh/6 + \genh/24);
	\node[anchor = east] at (#1, \genh/6) {\scalebox{\scale}{#4}};
	\node at (#1,\genh/3 + 0.5) {\scalebox{\scale}{#2}};
	\node at (#1+1,\genh/3+0.5) {\scalebox{\scale}{#3}};
}

\newcommand{\bboxcup}[4]{
	\draw[usual] (#1,\genh/3) \dr (#1+0.5,\genh/3-\genh/3) \ru (#1+1,\genh/3);
	\filldraw[usual, draw=black, fill = boxfill] (#1+1-\genh/24, \genh/3 - \genh/6 - \genh/24) rectangle (#1+1+\genh/24,\genh/3 -  \genh/6 + \genh/24);
	\node at (#1,\genh/3 + 0.5) {\scalebox{\scale}{#2}};
	\node at (#1+1,\genh/3+0.5) {\scalebox{\scale}{#3}};
	\node[anchor = east] at (#1+1, \genh/3 - \genh/6) {\scalebox{\scale}{#4}};
}

\newcommand{\capbox}[4]{
	\draw[usual] (#1,0) \ur (#1+.5,\genh/3) \rd (#1+1,0);
	\filldraw[usual, draw=black, fill = boxfill] (#1-\genh/24, \genh/6 - \genh/24) rectangle (#1+\genh/24, \genh/6 + \genh/24);
	\node[anchor = east] at (#1, \genh/6)  {\scalebox{\scale}{#4}};
	\node at (#1,-.5) {\scalebox{\scale}{#2}};
	\node at (#1+1,-.5) {\scalebox{\scale}{#3}};
}

\newcommand{\capbbox}[4]{
	\draw[usual] (#1,0) \ur (#1+.5,\genh/3) \rd (#1+1,0);
	\filldraw[usual, draw=black, fill = boxfill] (#1+1-\genh/24, \genh/6 - \genh/24) rectangle (#1+1+\genh/24, \genh/6 + \genh/24);
	\node at (#1,-.5) {\scalebox{\scale}{#2}};
	\node at (#1+1,-.5) {\scalebox{\scale}{#3}};
	\node[anchor = east] at (#1+1, \genh/6) {\scalebox{\scale}{#4}};
}

\newcommand{\subt}[2]{
	\draw[usual] (#1,0) to (#1,\genh/3);
	\node at (#1,\genh/3+ .5) {\scalebox{\scale}{#2}};
} 

\newcommand{\sub}[2]{
	\draw[usual] (#1,0) to (#1,\genh/3);
	\node at (#1,-.5) {\scalebox{\scale}{#2}};
} 

\newcommand{\cupb}[3]{
	\draw[usual] (#1,\genh/3) \dr (#1+.5,0) \ru (#1+1,\genh/3);
	\node at (#1,-.5) {\scalebox{\scale}{#2}};
	\node at (#1+1,-.5) {\scalebox{\scale}{#3}};
}

\newcommand{\capp}[3]{
	\draw[usual] (#1,0) \ur (#1+.5,\genh/3) \rd (#1+1,0);
	\node at (#1,-.5) {\scalebox{\scale}{#2}};
	\node at (#1+1,-.5) {\scalebox{\scale}{#3}};
}

\begin{document}

\title[The $\mathbb{Z}_{N}$ clock model and the coupled Temperley-Lieb algebra]{The planar parafermion algebra: The $\mathbb{Z}_{N}$ clock model and the coupled Temperley-Lieb algebra}

\author{Remy Adderton and Murray T. Batchelor}

\address{Mathematical Sciences Institute, 
The Australian National University, Canberra ACT 2601, Australia}


\begin{abstract}
The Hamiltonian of the $N$-state clock model is written in terms of a coupled Temperley-Lieb (TL) algebra defined by $N-1$ types of TL generators. 
This generalizes a previous result for $N=3$ obtained by J. F. Fjelstad and T. M\r{a}nsson [J. Phys. A {\bf 45} (2012) 155208]. 
The $\mathbb{Z}_{N}$-symmetric clock chain Hamiltonian expressed in terms of the coupled TL algebra generalizes the well known correspondence between the $N$-state Potts model and the TL algebra. 
The algebra admits a pictorial description in terms of a planar algebra involving parafermionic operators attached to $n$ strands.
%
%
A key ingredient in the resolution of diagrams is the string Fourier transform. The pictorial presentation also allows a description of the Hilbert space. 
We also give a pictorial description of the representation related to the staggered XX spin chain.
Just as the pictorial representation of the TL algebra has proven to be particularly useful in providing a visual and intuitive way to understand and manipulate algebraic expressions, 
it is anticipated that the pictorial representation of the coupled TL algebra  
may lead to further progress in understanding various aspects of the $\mathbb{Z}_{N}$ clock model, including the superintegrable chiral Potts model. 
\end{abstract}
\maketitle

\section{Introduction}

$\mathbb{Z}_N$ parafermions play a central role in the construction of a range of fundamental $N$-state interaction models in statistical and condensed matter physics with underlying $\mathbb{Z}_N$ symmetry. 
There has been a revival of interest in $\mathbb{Z}_N$ parafermion models in the context of parafermionic edge zero modes and topological phases~\cite{F,AF}.
The planar para algebra introduced by Jaffe and Liu~\cite{JaffeLiu} arises naturally as a planar algebra from combining planar algebras with $\mathbb{Z}_N$ para symmetry in physics.
The planar parafermion algebra~\cite{JaffeLiu} is used to show a horizontal reflection positivity property of a zero-graded Hamiltonian in the planar parafermion algebra. 
The planar parafermion algebra has also been used to develop a pictorial approach to quantum information where the string Fourier transform plays a role in creating states with maximal entanglement entropy \cite{JaffeHolo}.

The aim of this article is to generalize the role of the Temperley-Lieb (TL) algebra~\cite{TL,J83} in the Ising and $N$-state Potts model to that of the general $\mathbb{Z}_{N}$ clock model, including the superintegrable chiral Potts model (SICP)~\cite{HKN,vGR,McCoy} as a special case.
The algebraic connection between the 3-state SICP chain and two coupled copies of the TL algebra was established by Fjelstad and M\r{a}nsson~\cite{FM}. 
Here we recast the general $\mathbb{Z}_{N}$ clock model, including the SICP case, in a presentation in terms of either an $N-1$ or $N$ coupled TL algebra. 
A pictorial representation of this coupled algebra was given for the $N=3$ case which involves a generalisation of the pictorial presentation of the TL algebra to include a pole around which loops can become entangled~\cite{ABW}. However, in that case necessary far-apart commutation of all generators is not always satisfied. In this article we provide the correct pictorial description for general $N$. 
The key ingredient is the diagrammatic language of Jaffe and Liu's planar parafermion algebra, which naturally describes the general coupled TL algebra. In particular the string Fourier transform defines rotations in the algebra. 
We also give a diagrammatic description of the representation related to the staggered XX (sXX) spin chain~\cite{KT,PerkXX} discussed by Fjelstad and M\r{a}nsson~\cite{FM}.
Here rotations of the generators also play a key role in the pictorial description of the cubic relations in the algebra. 
The generators of the sXX representation are connected to those of a chromatic algebra, related to an invariant of trivalent planar tangles~\cite{EckRydberg}. 

We turn now to two of the key ingredients necessary for this work.

\subsection{ $\mathbb{Z}_{N}$ Clock Model}

The $\mathbb{Z}_{N}$ clock spin chain Hamiltonian is defined on a chain of length $L$ by 
\begin{equation}
	\label{clock1}
	H_{\mathrm{N}} = - \lambda \sum_{j=1}^{L} \sum_{m=1}^{N-1} \alpha_{m} (\tau_{j})^{m} -  \sum_{j=1}^{L-1} \sum_{m=1}^{N-1} \bar{\alpha}_{m} (\sigma_{j}^{\dagger} \sigma_{j+1})^{m}.
\end{equation}
The parameter $\lambda \in \mathbb{R}$ is a temperature-like coupling. 
The Hamiltonian is Hermitian when the coefficients $\alpha_{m}, \bar{\alpha}_{m} \in \mathbb{C}$ satisfy the conditions 
\begin{equation}
	\alpha_{m}^{*} = \alpha_{N-m}, \quad \bar{\alpha}_{m}^{*} = \bar{\alpha}_{N-m}.
\end{equation}
The shift and clock operators $\tau_j$ and $\sigma_j$ acting at site $j$ satisfy the relations 
\begin{equation}
\label{Paulirelations}
\tau_j^\dagger = \tau_j^{N-1}, \qquad \sigma_j^\dagger = \sigma_j^{N-1}, \qquad \sigma_j \tau_j = \omega \, \tau_j \sigma_j, \quad \omega = e^{ 2\pi \mathrm{i}/N},
\end{equation}
with $\tau_j^N = \sigma_j^N =1$, where $\dagger$ denotes the conjugate transpose.
In terms of matrices, 
\begin{eqnarray}
\tau_j &=& 1 \otimes 1 \otimes  \cdots \otimes 1 \otimes \tau \otimes 1 \otimes \cdots \otimes 1,\\
\sigma_j &=& 1 \otimes 1 \otimes  \cdots \otimes 1 \otimes \sigma \otimes 1 \otimes \cdots \otimes 1,
\end{eqnarray}
where $1$ is the $N \times N$ identity matrix and the shift and clock matricies are 
\begin{equation}
	\tau = 
	\left( \begin{array}{ccccccc}
		0 & 0 & 0 & \ldots & 0 & 1\\ 
		1 & 0 & 0 & \ldots & 0 & 0\\ 
		0 & 1 & 0 & \dots & 0 & 0\\
		\vdots & \vdots & \vdots & & \vdots & \vdots\\
		0 & 0 & 0 & \ldots & 1 & 0
	\end{array} \right), 
	\quad 
	\sigma = 
	\left( \begin{array}{ccccccc}
		1 & 0 & 0 & \ldots & 0 & 0\\ 
		0 & \omega & 0 & \ldots & 0 & 0\\ 
		0 & 0 & \omega^2 & \dots & 0 & 0\\
		\vdots &\vdots & \vdots & & \vdots & \vdots\\
		0 & 0 & 0 & \ldots & 0 & \omega^{N-1}
	\end{array} \right),
\end{equation}
are generalized Pauli matrices.
The Hamiltonian (\ref{clock1}) is invariant under the $\mathbb{Z}_{N}$ symmetry of increasing each spin, $\sigma_{j} \rightarrow \omega \sigma_{j}$, with corresponding symmetry generator
\begin{equation}
	\label{spinflipgenerator}
	R= \prod_{j=1}^{L} \tau_{j},
\end{equation}
satisfying $R^{N} = 1$.

%
The Hamiltonian (\ref{clock1}) for coefficients $\alpha_{m} = \bar{\alpha}_{m} = 1$ reduces to the quantum version of the $N$-state Potts model~\cite{qPotts}. 
The integrable chiral Potts model~\cite{McCoy, Perk} is defined when the coefficients $ \alpha_{m} , \bar{\alpha}_{m} $ are parametrized by two angles $\phi, \bar{\phi}$ as 
\begin{equation}\label{coefficientsintro}
	\alpha_{m} = \frac{e^{\mathrm{i} (2m -N) \phi /N }}{\sin m \pi /N}, \quad  \bar{\alpha}_{m} = \frac{e^{\mathrm{i} (2m -N) \bar{\phi} /N }}{\sin m \pi /N}.
\end{equation}
At $\phi=\bar{\phi}=0$ the model reduces to the Fateev-Zamolodchikov (FZ) model~\cite{FZ} 
\begin{equation}
	H_{\mathrm{FZ}} = -\sum_{j=1}^{L} \sum_{m=1}^{N-1} \frac{1}{\sin( m \pi /N)} \left( \lambda  \tau_{j}^{m} +\sigma_{j}^{m} \sigma_{j+1}^{-m} \right),
\end{equation}
which is equivalent to the Potts model for $N=3$. 
Another special case is when $\phi=\bar{\phi}=\pi/2$, corresponding to the $N$-state superintegrable chiral Potts (SICP) model, originating in discoveries by Howes, Kadanoff and den Nijs~\cite{HKN} and by von Gehlen and Rittenberg~\cite{vGR}. 
The SICP model is defined by the Hamiltonian~\cite{vGR,McCoy} 
\begin{equation}
H_{\mathrm{SICP}} = - \sum_{j=1}^L \sum_{m=1}^{N-1} \frac{2}{1-\omega^{-m}} ( \lambda \, \tau_j^m + (\sigma_j \sigma_{j+1}^\dagger)^m ) .
\label{SICP}
\end{equation}
The model defined by (\ref{SICP}) possesses additional symmetry generated by the Onsager algebra, owing to the Dolan-Grady condition~\cite{DG} being satisfied, beyond 
an infinite number of commuting conserved charges.
For this reason it is called superintegrable. 
The Onsager algebra plays a key role in 
solving the SICP chain for periodic boundary conditions~\cite{D90, D91}. 
We focus here particularly on the case of open boundary conditions, which are obtained by dropping the terms $(\sigma_L \, \sigma_{L+1}^\dagger )^m,$ with $m=1,2, \ldots, N-1$.

\subsection{The Temperley-Lieb Algebra}
The other main ingredient for the present work is the TL algebra~\cite{TL}, also known as the Temperley-Lieb-Jones algebra~\cite{J83}, 
which has enjoyed far reaching applications in both mathematics and physics. 
For each $n \in \mathbb{N}$ the TL algebra $\mathrm{TL}_{n}(q)$ is the unital associative algebra $\langle e_{i} | i \in \{1,\ldots,n-1\} \rangle$ subject to the relations
\begin{equation} 
 \label{tlrels}
e_{i}^{2}=(q+ q^{-1}) e_{i},\quad
e_{i}e_{j} = e_{j} e_{i},\quad e_{i}e_{i\pm 1}e_{i} =e_{i}, \quad |i-j|>1.
\end{equation}
The TL algebra underpins a number of key models in statistical mechanics \cite{Baxter, Martin}. 
For example the spin-1/2 XXZ and $N$-state Potts chains can be written in terms of generators $e_j$ satisfying the TL algebra relations, from which their TL equivalence is established \cite{ABB,Nichols2006}
\begin{equation}
\label{tlham}
H_{\mathrm{TL}} = -\sum_{j=1}^{L} e_{j}.
\end{equation}
Beyond the known representations in terms of spin operators, the TL algebra is arguably at its most powerful in the pictorial representation \cite{Martin, K, deGier}, 
with loop value $ \delta = q+q^{-1}$.
Various other generalisations of the TL algebra are known, e.g., multi-coloured TL algebras \cite{GP,BJ,GM}. 
The Fuss-Catalan algebra as well as the BMW algebra have been shown to be Yang-Baxter integrable, along with the Liu algebra \cite{Poncini2023}.
Here we give the pictorial representation for a coupled TL algebra of direct relevance to the $N$-state SICP model.

\section{The Coupled Temperley-Lieb Algebra}

For $n \in \mathbb{N}$, $N \ge 2$, the coupled TL algebra $\mathrm{cTL}_{n}(\sqrt{N})$ is the unital associative algebra with presentation $\braket{e_{i}^{(0)}, \ldots, e_{i}^{(N-1)} | i \in \{1,\ldots,n-1\}}$ subject to the relations 
\begin{eqnarray}
~~~~~~ e_{i}^{(k)}e_{i}^{(l)} &=& \delta_{k,l} \sqrt{N} e_{i}^{(k)} \label{ctlrel1}\\
~~~~~~ e_{i}^{(k)}e_{j}^{(l)} &=&e_{j}^{(l)} 	e_{i}^{(k)}, \quad |i-j| > 1 \label{ctlrel2}\\
e_{i}^{(k)}e_{i\pm 1 } ^{(l)}e_{i}^{(m)} &=& \frac{1}{\sqrt{N}} \sum_{p=1}^{N} \omega^{\mp (p-l)(k-m) } e_{i \pm 1}^{(p)} e_{i}^{(m)}\label{ctlrel3}\\
&=& \frac{1}{\sqrt{N}} \sum_{p=1}^{N} \omega^{\pm (p-l)(k-m) } e_{i}^{(k)}e_{i \pm 1}^{(p)}, \label{ctlrel4}\\
~~~~~~~~~~~~~~ 1 &=& \frac{1}{\sqrt{N}} \sum_{k=0}^{N-1} e_{i}^{(k)},  \label{identitygen}
\end{eqnarray}
for $k,l,m = 1, \ldots, N$. Here $\omega = e^{ 2\pi \mathrm{i}/N}$. 
The coupled TL algebra admits a natural presentation in terms of parafermion operators $c_{1}, \ldots, c_{n}$ satisfying
\begin{equation}
\label{parafermionrelations}
c_{i}^{N} =1, \quad c_{i}^{\dagger} = c_{i}^{N-1}, \quad c_{i} c_{j}  = \omega c_{j} c_{i}, \quad \mathrm{for~} i<j.
\end{equation}
For $N=2$ (\ref{parafermionrelations}) reduces to the well known anti-commutation relations for free fermions.
For an $L$ site Hilbert space $(\mathbb{C}^{N})^{\otimes L}$ we may define $2L$ parafermion operators $c_{1},\ldots, c_{2L}$ via the generalization of the Jordan-Wigner transformation known as the Fradkin-Kadanoff transformation \cite{FradkinKadanoff} 
\begin{equation}
\label{FKoperators}
c_{2i-1} = \bigg(\prod_{k=1}^{i-1}  \tau_{k} \bigg) \sigma_{i}, \quad c_{2i} = \omega^{\frac{N-1}{2}} \bigg(\prod_{k=1}^{i-1}  \tau_{k} \bigg) \sigma_{i} \tau_{i}.
\end{equation}
These operators satisfy (\ref{parafermionrelations}). A representation $\phi(e_{i}^{(k)})$ for $ k \in \mathbb{Z}_{N}$, of $\mathrm{cTL}_{2L}(\sqrt{N})$ is given by
\begin{eqnarray}
~~\phi(e_{2i-1}^{(k)}) &=& \frac{1}{\sqrt{N}} \sum_{m=1}^{N} (\omega^{\frac{2k -N +1}{2}} c_{2i-1}^{\dagger}c_{2i})^{m},\label{ctl1}\\
 	\quad \phi(e_{2i}^{(k)}) &=& \frac{1}{\sqrt{N}} \sum_{m=1}^{N} (\omega^{\frac{2k -N +1}{2}} c_{2i}c_{2i+1}^{\dagger})^{m},\label{ctl2}
\end{eqnarray}
i.e.,
\begin{eqnarray}
~~\phi(e_{2i-1}^{(k)}) &=& \frac{1}{\sqrt{N}} \sum_{m=1}^{N}  \omega^{\frac{m(m +2k-N)}{2}}  c_{2i-1}^{-m}c_{2i}^{m},\label{cctl1}\\
 	\quad \phi(e_{2i}^{(k)}) &=& \frac{1}{\sqrt{N}} \sum_{m=1}^{N} \omega^{\frac{m(m +2k-N)}{2}} c_{2i}^{m}c_{2i+1}^{-m}.\label{cctl2}
\end{eqnarray}
That is for each $k \in \mathbb{Z}_{N}$ the operators (\ref{ctl1})-(\ref{ctl2}) satisfy the relations of the usual TL algebra (\ref{tlrels}) and together satisfy the relations of the coupled TL algebra (\ref{ctlrel1})-(\ref{identitygen}). 
For ease of notation we denote $e_{i}^{(k)} = \phi(e_{i}^{(k)})$.
There exist additional relations between $\mathrm{cTL}_{n}(\sqrt{N})$ generators (\ref{ctl1})-(\ref{ctl2}) and the parafermion operators (\ref{FKoperators})
\begin{eqnarray} \label{paractlrelations}
~~~~~~ e_{2i-1}^{(k)}  &=& c_{2i-1}^{k} e_{2i-1}^{(0)} c_{2i-1}^{-k} = c_{2i}^{k}e_{2i-1}^{(0)} c_{2i}^{-k}, \\
~~~~~~ e_{2i}^{(k)}  &=& c_{2i}^{-k} e_{2i}^{(0)} c_{2i}^{k} = c_{2i+1}^{-k} e_{2i}^{(0)} c_{2i+1}^{k}.
\end{eqnarray}
Note the shift and spin difference terms of (\ref{clock1}) may be written as 
\begin{equation}
	\tau_{i} = \omega^{-\left(\frac{N-1}{2}\right)} c_{2i-1}^{\dagger} c_{2i}, \quad \sigma_{i} \sigma_{i+1}^{\dagger} =  \omega^{-\left(\frac{N-1}{2}\right)}  c_{2i} c_{2i+1}^{\dagger}.
\end{equation}

We may denote the representation (\ref{ctl1})-(\ref{cctl2}) on $(\mathbb{C}^{N})^{\otimes L}$ in terms of the generalized Pauli matrices as
\begin{equation}
	e_{2i-1}^{(k)} = \frac{1}{\sqrt{N}} \sum_{m=1}^{N} (\omega^{k} \tau_{i})^{m}, \quad 	e_{2i}^{(k)} = \frac{1}{\sqrt{N}} \sum_{m=1}^{N} (\omega^{k} \sigma_{i} \sigma_{i+1}^{\dagger})^{m},
\end{equation}
with loop value $\sqrt{N}$. 
%
%
%
%
The $\mathbb{Z}_{N}$ clock model Hamiltonian (\ref{clock1}) may be written 
\begin{equation} 
	\label{clockTL}
	H_{N} = -\lambda \sum_{j=1}^{L} \sum_{k=1}^{N} \hat{\bar{\alpha}}_{k} e_{2j-1}^{(k)}  - \sum_{j=1}^{L-1} \sum_{k=1}^{N} \hat{\alpha}_{k} e_{2j}^{(k)}, 
\end{equation}
with coefficients $\hat{\bar{\alpha}}_{k}, \hat{\alpha}_{k}$ given by 
\begin{eqnarray}
	\label{tlcoefficients}
	\hat{\bar{\alpha}}_{k} &= \frac{1}{\sqrt{N}} \sum_{m=1}^{N-1} \bar{\alpha}_{m} \omega^{-km},\quad \hat{\alpha}_{k}= \frac{1}{\sqrt{N}} \sum_{m=1}^{N-1} \alpha_{m} \omega^{-km}. 
\end{eqnarray}
In the chiral Potts model $\hat{\bar{\alpha}}_{k}, \hat{\alpha}_{k}$ are given by 
\begin{eqnarray}
	\hat{\bar{\alpha}}_{k}= \frac{1}{\sqrt{N}} \sum_{m=1}^{N-1} \frac{e^{\mathrm{i} (2m -N) \bar{\phi} /N }}{\sin m \pi /N}\omega^{-km}, \quad 
	\hat{\alpha}_{k}= \frac{1}{\sqrt{N}} \sum_{m=1}^{N-1} \frac{e^{\mathrm{i} (2m -N) \phi /N }}{\sin m \pi /N}\omega^{-km}. 
\end{eqnarray}

It is useful to write a Hamiltonian in terms of a presentation of $\mathrm{cTL}_{n}(\sqrt{N})$, which generalizes the single generator TL case of the Potts model. 
Due to the identity relation we may instead express the $\mathbb{Z}_{N}$ Hamiltonian up to an overall identity term and normalization in terms of the presentation involving $ e_{i}^{(k)}$ for $k =0,1, \ldots, N-2$.
In this presentation we choose to omit the $e^{(N-1)}_{i}$ generator and denote $e_{i} = e_{i}^{(0)}$.
It is thus natural to work in the $\mathrm{cTL}_{n}(\sqrt{N})$ presentation given by $\braket{1, e_{i}^{(0)}, \ldots, e_{i}^{(N-2)} | i \in \{1,\ldots,n-1\}}$ satisfying (\ref{ctlrel1})-(\ref{ctlrel2}) and the cubic relations
\begin{eqnarray}
\label{cubicrel1}
    e_{i}^{(k)}e_{i\pm 1 } ^{(l)}e_{i}^{(m)} &=& \frac{1}{\sqrt{N}} \sum_{p=0}^{N-2 } ( \omega^{\mp (p-l)(k-m) } - \omega^{\pm (l+1) (k-m)} ) e_{i \pm 1}^{(p)} e_{i}^{(m)} \label{sub3}\\
    &\phantom{=}& + \omega^{\pm (l+1) (k-m)} e_{i}^{(m)}, \nonumber
\end{eqnarray}
\begin{eqnarray}
\label{cubicrel2}
 e_{i}^{(k)}e_{i\pm 1 } ^{(l)}e_{i}^{(m)} &=& \frac{1}{\sqrt{N}} \sum_{p=0}^{N-2} ( \omega^{\pm (p-l)(k-m) } - \omega^{\mp (l+1) (k-m)} )  e_{i}^{(k)} e_{i \pm 1}^{(p)} \label{sub4}\\
 &\phantom{=}& + \omega^{\mp (l+1) (k-m)}e_{i}^{(k)}. \nonumber
\end{eqnarray}

Equivalently, we may write the clock and shift matrices in terms of the identity and $(N-1)$ $\mathrm{cTL}_{n} $ generators as 
\begin{eqnarray}
	~~~~~~ (\tau_j)^{m} &=& \omega^{m} + \frac{1}{\sqrt N} \sum_{k=0}^{N-2} (\omega^{-k m} - \omega^{m})  e_{2j-1}^{(k)}, \label{ctlinv1}\\
	(\sigma_j \sigma_{j+1}^\dagger)^{m} &=& \omega^{m} + \frac{1}{\sqrt N} \sum_{k=0}^{N-2} (\omega^{-k m} - \omega^{m})  e_{2j}^{(k)}. \label{ctlinv2}
\end{eqnarray}
For open boundaries the $\mathbb{Z}_{N}$ clock model Hamiltonian becomes
\begin{eqnarray}
	\label{clockTLsub}
	H_{N} = &-\lambda \sum_{j=1}^{L} \sum_{k=0}^{N-2} ( \hat{\bar{\alpha}}_{k}  - \hat{\bar{\alpha}}_{-1} )  e_{2j-1}^{(k)}  - \sum_{j=1}^{L-1} \sum_{k=0}^{N-2} ( \hat{\alpha}_{k} - \hat{\alpha}_{-1})e_{2j}^{(k)}\\
	&- \sqrt{N} L \lambda \hat{\bar{\alpha}}_{-1}  - \sqrt{N} (L-1) \hat{\alpha}_{-1}.\nonumber
\end{eqnarray}
The coefficients (\ref{tlcoefficients}) satisfy $\hat{\alpha}_{N-1} = \hat{\alpha}_{-1}$ and $\hat{\bar{\alpha}}_{N-1} = \hat{\bar{\alpha}}_{-1}$. Similarly, for periodic boundaries including an additional generator $ e_{2L}^{(k)}$, the algebra becomes infinite-dimensional and the Hamiltonain is given by 
\begin{eqnarray}
	\label{clockTL2}
	H_{N} = &-\lambda \sum_{j=1}^{L} \sum_{k=0}^{N-2} ( \hat{\bar{\alpha}}_{k}  - \hat{\bar{\alpha}}_{-1} )  e_{2j-1}^{(k)}  - \sum_{j=1}^{L} \sum_{k=0}^{N-2} ( \hat{\alpha}_{k} - \hat{\alpha}_{-1})e_{2j}^{(k)}\\
	&- \sqrt{N} L (\lambda \hat{\bar{\alpha}}_{-1} +\hat{\alpha}_{-1}). \nonumber
\end{eqnarray}
In the self dual case, with $\hat{\bar{\alpha}}_{k} = \hat{\alpha}_{k}$, and at $\lambda = 1$ we obtain
\begin{eqnarray} 
	\label{clockselfdual}
	H_{N} = &-\sum_{j=1}^{2L-1} \sum_{k=0}^{N-2} ( \hat{\alpha}_{k}  - \hat{\alpha}_{-1} )  e_{j}^{(k)} - \sqrt{N} \hat{\alpha}_{-1} (2L-1).
\end{eqnarray}
In the $N=2$ case the Hamiltonian (\ref{clockTLsub}) reduces to that of the one-dimensional Ising model, involving a single TL generator $e_{i}=e^{(0)}_{i}$. For open boundaries (\ref{tlcoefficients}) gives 
\begin{equation} 
    \label{N=2case}
	H_{\mathrm{Ising}} = -\lambda \sqrt{2} \sum_{j=1}^{L}  e_{2j-1} - \sqrt{2} \sum_{j=1}^{L-1} e_{2j} + L(\lambda+1) -1.
\end{equation}
That is, at $\lambda=1$ the Hamiltonian is given by the TL Hamiltonian (\ref{tlham}), up to an overall constant and normalization.
Similarly in the $N$-state Potts model case  given by $\alpha_{m} = \bar{\alpha}_{m} = 1$, the coefficients are given by 
\begin{eqnarray}
	\hat{\alpha}_{k} =	\hat{\bar{\alpha}}_{k}= \frac{1}{\sqrt{N}} \sum_{m=1}^{N-1} \omega^{-km}.
\end{eqnarray}
The coefficients are all equal for $k \ge 1$ thus $( \hat{\alpha}_{k}  - \hat{\alpha}_{-1} )=0$ and (\ref{clockTLsub}) becomes 
\begin{eqnarray} 
\label{PottsCase}
	H_{N} = &-\lambda \sqrt{N} \sum_{j=1}^{L} e_{2j-1}^{(0)}  - \sqrt{N} \sum_{j=1}^{L-1} e_{2j}^{(0)} + L (\lambda+1)-1.
\end{eqnarray}

Similarly in the Fateev-Zamolodchikov case, $\phi = \bar{\phi} =0$, with Hamiltonian
\begin{equation}
	\label{fzham}
	H_{\mathrm{FZ}} = - \sum_{j=1}^{L} \sum_{m=1}^{N-1}\frac{1}{\sin{m \pi/N}} \left( \lambda \tau_{j}^{m} + \sigma_{j}^{m} \sigma_{j+1}^{-m} \right).
\end{equation}
The coefficients are given by 
\begin{eqnarray}
	\hat{\alpha}_{k} =	\hat{\bar{\alpha}}_{k}= \frac{1}{\sqrt{N}} \sum_{m=1}^{N-1} \frac{\omega^{-km}}{\sin m \pi /N} =   \frac{2\mathrm{i}}{\sqrt{N}} \sum_{m=1}^{N-1} \frac{\omega^{-m(k +1/2)}}{1 - \omega^{-m}}.
\end{eqnarray}
Note that $\hat{\alpha}_{1} = \hat{\alpha}_{-1} $ independent of $N$, hence the coefficient of $e^{(1)}_{i}$ vanishes in the FZ case and $N-2$ coupled Temperley-Lieb generators appear in the Hamiltonian
\begin{eqnarray} 
	\label{fzham2}
	H_{FZ} = &-\lambda \sum_{j=1}^{L} \sum_{k=0}^{N-2} ( \hat{\bar{\alpha}}_{k}  - \hat{\bar{\alpha}}_{1} )  e_{2j-1}^{(k)}  - \sum_{j=1}^{L-1} \sum_{k=0}^{N-2} ( \hat{\alpha}_{k} - \hat{\alpha}_{1})e_{2j}^{(k)}\\
	&- \sqrt{N} (L(\lambda +1)-1) \hat{\bar{\alpha}}_{1}.\nonumber
\end{eqnarray}
In the $N=3$ case, the Hamiltonian (\ref{fzham2}) describes the $3$-state Potts model Hamiltonian (\ref{PottsCase}) up to an overall normalization
\begin{eqnarray} 
	H_{\mathrm{FZ}} = &-\lambda  \sum_{j=1}^{L} 2 e_{2j-1}^{(0)} -  \sum_{j=1}^{L-1}  2 e_{2j}^{(0)} +\frac{2}{\sqrt{3}} (L(\lambda+1)-1).
\end{eqnarray}

For general $N$ the SICP Hamiltonian may be written as 
%
\begin{equation}
	H_{\mathrm{SICP}} = \frac{1}{\sqrt{N}} \sum_{j=1}^{L} \sum_{k=0}^{N-1} (\frac{1}{2}(N-1) + k)(\lambda e_{2j-1}^{(k)} +  e_{2j}^{(k)}),
\end{equation}
using Eq.~(2.18) of \cite{Baxter09} to simplify the coefficients $\hat{\alpha}_{m}, \hat{\bar{\alpha}}_{m}$.
As shown in \cite{FM} under the relabeling $\tau \rightarrow \omega \tau$ the $N=3$ SICP Hamiltonian may be expressed under periodic boundary conditions as
\begin{equation}
    H_{\mathrm{SICP}} = - \frac{4}{\sqrt{3}} \sum_{j=1}^{L} [\lambda (e_{2j-1}^{(1)} -  e_{2j-1}^{(2)} ) + (e_{2j}^{(1)} -  e_{2j}^{(2)} )].
\end{equation}

\section{Planar Parafermion Algebra}
In this section we follow the construction of \cite{JaffeLiu} in order to define a pictorial representation of $\mathrm{cTL}_{n}(\sqrt{N})$. 
For $n \in \mathbb{N}$ the planar parafermion (para) algebra $\mathrm{PF}_{n}$ is given by $\mathbb{Z}_{N}$-labelled 1-boxes on $n$ strands \cite{JaffeLiu}.
 The basis elements of the planar parafermion algebra are the labelled 1-boxes of the form
\begin{equation}
	\begin{tikzpicture}[anchorbase, scale=0.5]
		\su{0}{1};
		\node at (1, \genh/2){$\cdots$} ;
		\su{2}{};
		\kbox{3}{$i$}{$k$};
		\su{4}{};
		\node at (5, \genh/2){$\cdots$} ;
		\su{6}{$n$};
	\end{tikzpicture},
\end{equation}
with $k \in \mathbb{Z}_{N}$.
The $k$-labelled 1-boxes satisfies a `para-isotopy' relation
\begin{equation}
\label{paraisotopy}
	\begin{tikzpicture}[anchorbase, scale=0.75]
		\su{0}{1};
		\node at (1, \genh/2){$\cdots$} ;
		\kboxtl{2}{$i$}{$k$};
		\node at (3, \genh/2) {$\cdots$};
		\kboxbl{4}{$j$}{$l$};
		\node at (5, \genh/2){$\cdots$};
		\su{6}{$n$};
	\end{tikzpicture}
	=\omega^{kl}
	\begin{tikzpicture}[anchorbase, scale=0.75]
		\su{0}{1};
		\node at (1, \genh/2){$\cdots$} ;
		\kboxbl{2}{$i$}{$k$};
		\node at (3, \genh/2) {$\cdots$};
		\kboxtl{4}{$j$}{$l$};
		\node at (5, \genh/2){$\cdots$};
		\su{6}{$n$};
	\end{tikzpicture},
\end{equation}
for $\omega = e^{2 \pi \mathrm{i}/N}$. 
The $k$-labelled 1-boxes provide a representation of $\mathbb{Z}_{N}$ on a single strand, with labels treated mod $N$ and contractable loops take value $\sqrt{N}$ and $k$-valued loops take value zero for $k\neq 0$, i.e.,
\begin{equation}
\label{ZNproperty}
	\begin{tikzpicture}[anchorbase, scale=0.75]
		\kboxd{2}{$i$}{$k$}{$l$};
	\end{tikzpicture}
	= 
	\begin{tikzpicture}[anchorbase, scale=0.75]
		\kbox{2}{$i$}{$k+l$};
	\end{tikzpicture},\qquad \qquad
      \begin{tikzpicture}[anchorbase,scale=0.75]
		\draw[usual] (0,0) circle (\genh/6);
		\filldraw[usual, fill = boxfill] (-\genh/6 - \genh/24, -\genh/24) rectangle (-\genh/6 + \genh/24, \genh/24);
		\node[anchor=east] at (-\genh/6, 0){\scalebox{\scale}{$k$}};
	\end{tikzpicture}
	= \sqrt{N}\delta_{k,0} ,
\end{equation}
with $k$ $\mathrm{mod~} N$. 

In the pictorial approach it is useful to define a twisted tensor product on $PF_{n}$, given by placing 1-boxes at the same horizontal level
\begin{equation} 
\label{tproduct}
	\raisebox{-.25cm}{
		\begin{tikzpicture}[anchorbase,scale=0.75]
			\kboxhd{0}{$ i $}{$ k $ };
			\kboxhd{1}{$ j $}{$ l $ };
	\end{tikzpicture}}
    := \omega^{-\frac{kl}{2}} 
    	\raisebox{-.25cm}{
		\begin{tikzpicture}[anchorbase,scale=0.75]
			\kboxh{1}{$ j $}{$ l $ };
			\sub{0}{$ i $};
			\begin{scope}[shift={(0,\genh/3)}]
				\kboxh{0}{~}{$ k $ };
				\sub{1}{~};
			\end{scope}
	\end{tikzpicture}}
    =
    \omega^{\frac{kl}{2}} 
	\raisebox{-.25cm}{
		\begin{tikzpicture}[anchorbase,scale=0.75]
			\kboxh{0}{$ i $}{$ k $ };
			\sub{1}{$ j $};
			\begin{scope}[shift={(0,\genh/3)}]
				\kboxh{1}{}{ $ l $ };
				\sub{0}{~};
			\end{scope}
	\end{tikzpicture}}, \quad i < j. 
\end{equation}
The representation (\ref{ctl1})-(\ref{ctl2}) may then be written as 
\begin{equation}
	\begin{tikzpicture}[anchorbase, scale=0.75]
		\smallboxcupcapbox{0}{$2i-1$}{$2i$}{$k$}{$-k$};
	\end{tikzpicture}
	= \frac{1}{\sqrt{N}} \sum_{n=1}^{N} \omega^{n(k - \frac{N}{2})}
	\begin{tikzpicture}[anchorbase, scale=0.75]
		\kboxhd{0}{$2i-1$}{$-n$};
		\kboxhd{1}{$2i$}{$n$}
	\end{tikzpicture}, 
    \label{ctltwisted1}
\end{equation}
\begin{equation}
\label{ctltwisted2}
	\begin{tikzpicture}[anchorbase, scale=0.75]
		\smallboxcupcapbox{0}{$2i$}{$2i+1$}{$-k$}{$k$};
	\end{tikzpicture}
	= \frac{1}{\sqrt{N}} \sum_{n=1}^{N}  \omega^{n(k - \frac{N}{2})}
	\begin{tikzpicture}[anchorbase, scale=0.75]
		\kboxhd{0}{$2i$}{$n$};
		\kboxhd{1}{$2i+1$}{$-n$}
	\end{tikzpicture}.
\end{equation}
Note that for $k=0$ the usual TL generator differs to that of \cite{JaffeLiu} by an additional $\omega^{-\frac{nN}{2}}$ factor in the sum of (\ref{ctltwisted1}) and (\ref{ctltwisted2}).
This defines $e^{(0)}_{i}$ as the $N$-state Potts representation of the Temperley-Lieb algebra. Diagrammatically it is represented by an unlabelled cup and cap in the parafermion planar algebra.
The identity operator in $PF_{n}$ can we written as a sum of graded cups and caps as 
\begin{equation}
\label{identitypictoral}
	\raisebox{-0.125cm}{
		\begin{tikzpicture}[anchorbase, scale=.75]
			\suhb{0}{};
			\suhb{1}{};
	\end{tikzpicture}}
	=  \frac{1}{\sqrt{N}}  \sum_{k=1}^{N} 
	\raisebox{-0.125cm}{
		\begin{tikzpicture}[anchorbase, scale=0.75]
			\smallboxcupcapbox{1}{~}{~}{$k$}{$-k$};
	\end{tikzpicture}}.
\end{equation}

There exist additional relations between operators (\ref{ctltwisted1})-(\ref{ctltwisted2}) and the parafermion operators, given by the following cup diagrams
\begin{eqnarray}
\label{cupboxes}
	\begin{tikzpicture}[anchorbase, scale =0.75]
	\boxcup{0}{$2i-1$}{$2i$}{$k$};
	\end{tikzpicture}
	&=&
\omega^{\frac{-(N -k)k}{2}}
	\begin{tikzpicture}[anchorbase, scale =0.75]
	\bboxcup{0}{$2i-1$}{$2i$}{$k$};
	\end{tikzpicture}, \quad
	\begin{tikzpicture}[anchorbase, scale =0.75]
		\boxcup{0}{$2i$}{$2i+1$}{$k$};
	\end{tikzpicture}
	=
	\omega^{\frac{(N +k)k}{2}}
	\begin{tikzpicture}[anchorbase, scale =0.75]
		\bboxcup{0}{$2i$}{$2i+1$}{$k$};
	\end{tikzpicture},
\end{eqnarray}
and similarly for cap diagrams 
\begin{eqnarray}
\label{capboxes}
	\begin{tikzpicture}[anchorbase, scale =0.75]
		\capbox{0}{$2i-1$}{$2i$}{$k$};
	\end{tikzpicture}
	&=&
	\omega^{\frac{-(N +k)k}{2}}
	\begin{tikzpicture}[anchorbase, scale =0.75]
		\capbbox{0}{$2i-1$}{$2i$}{$k$};
	\end{tikzpicture},\quad
	\begin{tikzpicture}[anchorbase, scale =0.75]
		\capbox{0}{$2i$}{$2i+1$}{$k$};
	\end{tikzpicture}
	=
	\omega^{\frac{(N -k)k}{2}} 
	\begin{tikzpicture}[anchorbase, scale =0.75]
		\capbbox{0}{$2i$}{$2i+1$}{$k$};
	\end{tikzpicture}.
\end{eqnarray}

The coupled Temperley-Lieb algebra introduced above admits a natural diagrammatic presentation in the planar parafermion algebra. 
For $N \ge 2$, $k \in \mathbb{Z}_{N}$ a representation of $\mathrm{cTL}_{n}(\sqrt{N})$ within the planar parafermion algebra is given by
\begin{eqnarray}
\label{ctlgen1}
	e_{2i-1}^{(k)} &:=&
	\raisebox{-0.25cm}{
	\begin{tikzpicture}[anchorbase, scale=0.75]
		\sutl{0}{1};
		\node at (1, 3*\genh/8){$\cdots$} ;
		\sutl{2}{~};
        \smallboxcupcapbox{3}{$2i-1$}{$2i$}{$k$}{$-k$};
		\sutl{5}{};
		\node at (6, 3*\genh/8){$\cdots$};
		\sutl{7}{$n$};
	\end{tikzpicture}},\\
	e_{2i}^{(k)} &:=&
	\raisebox{-0.25cm}{
	\begin{tikzpicture}[anchorbase, scale=0.75]
		\sutl{0}{1};
		\node at (1, 3*\genh/8){$\cdots$} ;
		\sutl{2}{~};
        \smallboxcupcapbox{3}{$2i$}{$2i+1$}{$-k$}{$k$};
		\sutl{5}{};
		\node at (6, 3*\genh/8){$\cdots$};
		\sutl{7}{$n$};
	\end{tikzpicture}}. \label{ctlgen2}
\end{eqnarray}
Here $n = 2L$.
The diagrams form a representation of $\mathrm{cTL}_{n}(\sqrt{N})$. The orthogonality relation (\ref{ctlrel1}) is satisfied through the $k$-loop relation
\begin{equation*}
	e_{2i-1}^{(k)} e_{2i-1}^{(l)} = 
	\begin{tikzpicture}[anchorbase, scale =0.5]
		\smallboxcupcapbox{0}{$i$}{$i+1$}{$l$}{$-l$};
		\begin{scope}[shift = {(0,5*\genh/6)}]
			\smallboxcupcapbox{0}{~}{~}{$k$}{$-k$};
		\end{scope}
	\end{tikzpicture}
	=
	\begin{tikzpicture}[anchorbase, scale =0.5]
		\draw[usual] (0,0) circle (\genh/6);
		\filldraw[usual, fill =boxfill] (-\genh/6 - \genh/24, -\genh/24) rectangle (-\genh/6 + \genh/24, \genh/24);
		\node[anchor= east] at (-\genh/6, 0) {\scalebox{\scale}{$l-k$}}; 
	\end{tikzpicture}
	\begin{tikzpicture}[anchorbase, scale =0.5]
		\smallboxcupcapbox{0}{$i$}{$i+1$}{$k$}{-$l$};
	\end{tikzpicture}
	=
	\sqrt{N} \delta_{l,k}
	\begin{tikzpicture}[anchorbase, scale =0.5]
		\smallboxcupcapbox{0}{$i$}{$i+1$}{$k$}{-$k$};
	\end{tikzpicture}.
\end{equation*}
And similarly for $e_{2i}^{(k)} e_{2i}^{(l)}$. The relations (\ref{paractlrelations}) are trivially satisfied. The relation (\ref{ctlrel2}) describes far-apart commutativity and follows from the para-isotopy relation (\ref{paraisotopy}) for the $\pm k$ labelled cups and caps in (\ref{ctlgen1})-(\ref{ctlgen2}).
A natural diagrammatic representation of the algebra (\ref{parafermionrelations}) is given by denoting $c_{i}^{k}$ as a $k$-labelled 1-box on the $i$-th strand.
The basis elements of the planar parafermion algebra are the labelled 1-boxes of the form
\begin{equation}
	c_{i}^{k} =
	\begin{tikzpicture}[anchorbase, scale=0.5]
		\su{0}{1};
		\node at (1, \genh/2){$\cdots$} ;
		\su{2}{};
		\kbox{3}{$i$}{$k$};
		\su{4}{};
		\node at (5, \genh/2){$\cdots$} ;
		\su{6}{$n$};
	\end{tikzpicture}.
\end{equation}
Multiplication is given by stacking diagrams with left-to-right becoming top-to-bottom
\begin{equation}
	c_{1}^{k_{1}} c_{2}^{k_{2}} \cdots c_{n}^{k_{n}} 
	=
	\raisebox{-.25cm}{
		\begin{tikzpicture}[anchorbase, scale =0.75]
			\kboxh{0}{}{$k_{1}$};
			\sub{1}{}{};
			\sub{3}{}{};
			\begin{scope}[shift={(0, - \genh/3)}]
				\kboxh{1}{}{$k_{2}$};
				\sub{0}{}{};
				\sub{3}{}{};
			\end{scope}
			\begin{scope}[shift={(0, -2*\genh/3)}]
				\kboxh{3}{$n$}{$k_{n}$};
				\sub{0}{$1$}{};
				\sub{1}{$2$}{};
				\node at (1.75, \genh/6) {$\cdots$};
			\end{scope}
	\end{tikzpicture}}.
\end{equation}
In the algebraic presentation (\ref{cupboxes})-(\ref{capboxes}) are given by
\begin{eqnarray}
c_{2i-1}^{k}  e_{2i-1}^{(0)} &=&   \omega^{-k(\frac{N-k}{2})}  c_{2i}^{k} e^{(0)}_{2i-1},\quad 
c_{2i}^{k} 	e_{2i}^{(0)} = \omega^{k(\frac{N+k}{2})} c_{2i+1}^{k} e^{(0)}_{2i},\\
e_{2i-1}^{(0)} c_{2i-1}^{k} &=&   \omega^{-k(\frac{N+k}{2})}  e^{(k)}_{2i-1} c_{2i}^{k},\quad
e_{2i}^{(0)} c_{2i}^{k} = \omega^{k(\frac{N-k}{2})}  e^{(0)}_{2i} c_{2i+1}^{k}.
\end{eqnarray}

\subsection{String Fourier Transform}
A key-ingredient in the cubic relations in the diagrammatic presentation is the relation between the generators of the coupled algebra and their one-strand or one-click rotations. In the planar parafermion algebra these rotations are described by a string Fourier transform (SFT) \cite{JaffeLiu}.
The SFT may be defined via the actions of inclusion and conditional expectation on the planar algebra. 
The conditional expectation is the map $\epsilon: PF_{n} \rightarrow PF_{n-1}$
%
\begin{equation}
	\epsilon(x)= \frac{1}{\delta}
	\raisebox{-0.25cm}{
		\begin{tikzpicture}[anchorbase, scale=0.5]
			\sub{0}{$1$}:
			\sub{1}{$2$};
			\sub{3}{$n-1$};
			\draw[usual, draw= black] (5,\genh/3) to (5, 3*\genh/3); 
			\draw[usual, draw= black] (4,\genh/3) to (4, 3*\genh/3); 
			\node at (2,\genh) {$\cdots$};
			\node at (2,\genh/3) {$\cdots$};
			\shortcup{4}{~}{~};
			\begin{scope}[shift={(0,\genh/3)}]
				\su{0}{~};
				\su{1}{~};
				\su{3}{~};
				\filldraw[usual, draw=black, fill = boxfill] (0-0.25, \genh/3 -0.5) rectangle (3 +1.25, \genh/3+0.5)node [pos=.5]{$x$};
				\filldraw[black] (0-0.25, \genh/3) circle (1pt);
			\end{scope}
			\begin{scope}[shift={(0,3*\genh/3)}]
				\sub{0}{~};
				\shortcap{4}{~};
			\end{scope}
	\end{tikzpicture}},
\end{equation}
and the left and right inclusion $\iota_{l,r}: PF_{n} \rightarrow  PF_{n+1} $ is defined by adding a strand to the left (resp. right) of an $n$-box. For example 
\begin{equation}
		\iota_{l} (x)= \frac{1}{\delta}
		\raisebox{-0.25cm}{
			\begin{tikzpicture}[anchorbase, scale=0.5]
				\sub{-1}{$1$}
				\sub{0}{$2$}:
				\sub{1}{$3$};
				\sub{3}{$n$};
				\sub{4}{};
                \node at (4.5,-.5) {\scalebox{\scale}{$n+1$}};
				\draw[usual, draw= black] (-1,0) to (-1, 3*\genh/3); 
				\draw[usual, draw= black] (4,0) to (4, 3*\genh/3); 
				\node at (2,\genh) {$\cdots$};
				\node at (2,\genh/3) {$\cdots$};
				\begin{scope}[shift={(0,\genh/3)}]
					\su{0}{~};
					\su{1}{~};
					\su{3}{~};
					\sub{4}{~};
					\sub{-1}{~};
					\filldraw[usual, draw=black, fill = boxfill] (0-0.25, \genh/3 -0.5) rectangle (3 +1.25, \genh/3+0.5)node [pos=.5]{$x$};
					\filldraw[black] (0-0.25, \genh/3) circle (1pt);
				\end{scope}
				\begin{scope}[shift={(0,3*\genh/3)}]
					\sub{0}{~};
					\sub{4}{~};
					\sub{-1}{~};
				\end{scope}
		\end{tikzpicture}},
\end{equation}
for an $n$-box $x\in PF_{n}$. Here the marked point on the left of the n-box defines the orientation and the strands may be labelled $1,\ldots, n$ from left-to-right along the bottom, and $n+1, \ldots, 2n$ from right-to-left on the top. Following the notation of Jaffe-Liu \cite{JaffeLiu}, the sting Fourier transform $\mathcal{F}_{s}: PF_{n} \rightarrow PF_{n}$ is defined as 
\begin{equation} \label{fsaction}
		\mathcal{F}_{s}(x)=
		\raisebox{-0.25cm}{
			\begin{tikzpicture}[anchorbase, scale=0.5]
				\sub{1}{~};
				\sub{3}{~};
				\sub{4}{~};
				\draw[usual, draw= black] (5,0) to (5, 3*\genh/3); 
				\draw[usual, draw= black] (4,0) to (4, 3*\genh/3); 
				\shortcup{-1}{~}{~};
				\begin{scope}[shift={(0,\genh/3)}]
					\su{-1}{~};
					\su{0}{~};
					\su{1}{~};
					\su{3}{~};
					\sub{4}{~};
					\sub{5}{~};
					\filldraw[usual, draw=black, fill = boxfill] (0-0.25, \genh/3 -0.5) rectangle (3 +1.25, \genh/3+0.5)node [pos=.5]{$x$};
					\filldraw[black] (0-0.25, \genh/3) circle (1pt);
				\end{scope}
				\begin{scope}[shift={(0,3*\genh/3)}]
					\shortcap{4}{~}{~};
					\sub{-1}{~};
					\sub{0}{~};
				\end{scope}
		\end{tikzpicture}} = \delta \epsilon (\iota_{l}(x) e_{1} e_{2} \cdots e_{n}), 
\end{equation}
%
for an $n$-box $x \in PF_{n}$. The action (\ref{fsaction}) is also referred to as a 1-click rotation of the $m$-box. We may similarly define the inverse string Fourier transform $\mathcal{F}^{-1}_{s}: PF_{n} \rightarrow PF_{n}$ by $\mathcal{F}^{-1}_{s}(x) = \delta \epsilon (\iota_{l}(x) e_{n} e_{n-1} \cdots e_{1})$.
For $n=2$ the action of $\mathcal{F}_{s}$ (and $\mathcal{F}^{-1}_{s}$) amounts to a clockwise (resp. anti-clockwise) $\frac{\pi}{2}$-rotation
\begin{equation}
\label{sft}
	\mathcal{F}_{s} \left(
		\raisebox{-0.125cm}{
		\begin{tikzpicture}[anchorbase, scale=0.5]
			\sub{0}{~};
			\sub{1}{~};
			\begin{scope}[shift={(0, \genh/3)}]
				\sub{0}{~};
				\sub{1}{~};
				\twobox{0}{$x$};
			\end{scope}
			\begin{scope}[shift={(0, 2*\genh/3)}]
				\sub{0}{~};
				\sub{1}{~};
			\end{scope}
	\end{tikzpicture}}
	\right)
	=
	\raisebox{-0.125cm}{
	\begin{tikzpicture}[anchorbase, scale=0.5]
		\shortcup{-1}{~}{~};
		\sub{1}{~};
		\sub{2}{~};
		\begin{scope}[shift={(0,\genh/3)}]
			\sub{-1}{~};
			\sub{0}{~};
			\sub{1}{~};
			\twobox{0}{$x$};
			\sub{2}{~};
		\end{scope}
		\begin{scope}[shift={(0,2*\genh/3)}]
			\sub{-1}{~};
			\sub{0}{~};
			\shortcap{1}{~}{~};
		\end{scope}
	\end{tikzpicture}}, \quad 
	\mathcal{F}_{s}^{-1} \left(
	\raisebox{-0.125cm}{
		\begin{tikzpicture}[anchorbase, scale=0.5]
			\sub{0}{~};
			\sub{1}{~};
			\begin{scope}[shift={(0, \genh/3)}]
				\sub{0}{~};
				\sub{1}{~};
				\twobox{0}{$x$};
			\end{scope}
			\begin{scope}[shift={(0, 2*\genh/3)}]
				\sub{0}{~};
				\sub{1}{~};
			\end{scope}
	\end{tikzpicture}}
	\right)
	=
	\raisebox{-0.125cm}{
		\begin{tikzpicture}[anchorbase, scale=0.5]
			\shortcup{1}{~}{~};
			\sub{-1}{~};
			\sub{0}{~};
			\begin{scope}[shift={(0,\genh/3)}]
				\sub{-1}{~};
				\sub{0}{~};
				\sub{1}{~};
				\twobox{0}{$x$};
				\sub{2}{~};
			\end{scope}
			\begin{scope}[shift={(0,2*\genh/3)}]
				\sub{1}{~};
				\sub{2}{~};
				\shortcap{-1}{~}{~};
			\end{scope}
	\end{tikzpicture}}.
\end{equation}

The action of the SFT on the generators of $\mathrm{cTL}_{n}(\sqrt{N})$ is given by 
\begin{equation}
\label{sftgens}
	\mathcal{F}_{s}^{\pm 1} \big(e_{i}^{(k)}\big)= \frac{1}{\sqrt{N}} \sum_{m=1}^{N} \omega^{\mp mk} e_{i}^{(m)}, 
\end{equation}
for $i=1, \ldots, n-1$. In the $k=0$ case, the action of the one-string Fourier transform (and its inverse) maps $e_{i}^{(0})$ to the identity on two strands $1_{2}$, i.e. $\mathcal{F}_{s} \big(e_{i}^{(0)}\big) = 1_{2}$ and (\ref{sftgens}) reduces to the identity relation (\ref{identitypictoral}). In this case the action of the both the string Fourier transform and its inverse on an ungraded cup and cap are equivalent 
\begin{equation}
	\mathcal{F}_{s}^{\pm 1} (e^{(0)}_{i} ) = 1_{2}, \qquad \mathcal{F}_{s}^{\pm 1} (1_{2}) = (e^{(0)}_{i} ).
\end{equation}
For $k\ge 1 $ the string Fourier transform takes the cTL generators to a pair of parafermions. For $i$ odd the SFT action (\ref{sft}) gives
\begin{equation}
\mathcal{F}_{s}(e_{2i-1}^{(k)})
	=
	\raisebox{0cm}{
	\begin{tikzpicture}[anchorbase, scale=0.75]
		\cupb{0}{~}{~};
		\sub{2}{~};			
		\sub{3}{~};
		\begin{scope}[shift={(0,\genh/3)}]
			\su{0}{~};
			\boxcupcapbox{1}{~}{~}{$k$}{$-k$};
			\su{3}{~}{~};
		\end{scope}
		\begin{scope}[shift={(0,\genh + \genh/3)}]
			\sub{0}{~};
			\sub{1}{~};
			\capp{2}{~}{~};
		\end{scope}
	\end{tikzpicture}}
	=
	\omega^{\frac{k(N-k)}{2}}
	\begin{tikzpicture}[anchorbase, scale=0.75]
		\kboxh{0}{~}{$-k$};
		\sub{1}{};
		\begin{scope}[shift={(0,\genh/3)}]
		\kboxh{1}{~}{$k$};
		\sub{0}{};
	\end{scope}
	\end{tikzpicture}, 
\end{equation}
where the left strand is resolved via the isotopy relation (\ref{capboxes}) for a strand on an even site. By inserting the identity relation (\ref{identitypictoral}) and resolving the resulting diagram using the parafermion commutation relation (\ref{paraisotopy}) and relations (\ref{cupboxes}) give the result
\begin{equation}
\mathcal{F}_{s}(e_{2i-1}^{(k)})	=
	\omega^{\frac{k(N-k)}{2}} \frac{1}{\sqrt{N}} \sum_{m=0}^{N-1}
	\begin{tikzpicture}[anchorbase, scale =0.75]
		\kboxh{0}{~}{$-k$};
		\sub{1}{};
		\begin{scope}[shift={(0,\genh/3)}]
		\kboxh{1}{~}{$k$};
		\sub{0}{};
	\end{scope}
    \begin{scope}[shift={(0,-\genh)}]
			\boxcupcapbox{0}{~}{~}{$m$}{$-m$};
		\end{scope}
	\end{tikzpicture}
      \frac{1}{\sqrt{N}} \sum_{m=0}^{N-1}  \omega^{-mk}
	\begin{tikzpicture}[anchorbase, scale =0.75]
    \begin{scope}[shift={(0,0*\genh)}]
			\boxcupcapbox{0}{~}{~}{$m$}{$-m$};
		\end{scope}
	\end{tikzpicture}.
\end{equation}

For the inverse SFT $\mathcal{F}_{s}^{-1}$ (\ref{sft}) gives
\begin{equation}\label{ftinverse}
	\mathcal{F}_{s}^{-1} \left(e_{2i-1}^{(k)}\right)
 	= 	\omega^{\frac{k(N-k)}{2}}
	\begin{tikzpicture}[anchorbase, scale =0.75]
			\kboxh{1}{~}{$-k$};
			\sub{0}{};
		\begin{scope}[shift={(0,\genh/3)}]
			\kboxh{0}{~}{$k$};
			\sub{1}{};
		\end{scope}
	\end{tikzpicture}
        =
     \frac{1}{\sqrt{N}} \sum_{m=0}^{N-1}  \omega^{mk}
	\begin{tikzpicture}[anchorbase, scale =0.75]
    \begin{scope}[shift={(0,0*\genh)}]
			\boxcupcapbox{0}{~}{~}{$m$}{$-m$};
		\end{scope}
	\end{tikzpicture}.
\end{equation}
And similarly for the action on $e_{2i-1}^{(k)}$. 
The relations between the one strand rotations in the sense of \cite{JonesQuadraticTangles} of the cTL generators was also observed in \cite{PonciniPreprint}.
By the application of the string Fourier transform in the diagrammatic presentation, the cubic relations (\ref{ctlrel3})-(\ref{ctlrel4}) follow. For example, for $e_{2i-1}^{(k)} e_{2i}^{(l)} e_{2i-1}^{(m)}$ we have by para-isotopy
\begin{equation*}
	e_{2i-1}^{(k)} e_{2i}^{(l)} e_{2i-1}^{(m)} = 
	\begin{tikzpicture}[anchorbase, scale =0.75]
		\boxcupcapbox{0}{~}{~}{$m$}{$-m$};
		\su{2}{~};
		\begin{scope}[shift = {(0,\genh)}]
			\su{0}{~};
			\eboxcupcapbox{1}{~}{~}{$-l$}{$l$};
		\end{scope}
		\begin{scope}[shift = {(0,2*\genh)}]
			\su{2}{~};
			\boxcupcapbox{0}{~}{~}{$k$}{$-k$};
		\end{scope}
	\end{tikzpicture}
	=
	\begin{tikzpicture}[anchorbase, scale =0.75]
		\boxcupcapbox{0}{~}{~}{$m$}{$-m$};
		\su{2}{~};
		\begin{scope}[shift = {(0,\genh)}]
			\su{0}{~};
			\ebboxcupcapbox{1}{~}{~}{$-l$}{$l$};
		\end{scope}
		\begin{scope}[shift = {(0,2*\genh)}]
			\su{2}{~};
			\boxcupcapbox{0}{~}{~}{$k$}{$-k$};
		\end{scope}
	\end{tikzpicture}
	= 
	\omega^{kl} \omega^{-ml} 
	\begin{tikzpicture}[anchorbase, scale =0.75]
		\boxcupcapbox{0}{~}{~}{$m$}{$-m$};
		\kbox{2}{~}{$l$};
		\begin{scope}[shift = {(0,\genh)}]
			\su{0}{~};
			\cupcap{1}{~}{~};
		\end{scope}
		\begin{scope}[shift = {(0,2*\genh)}]
			\kbox{2}{~}{$-l$};
			\boxcupcapbox{0}{~}{~}{$k$}{$-k$};
		\end{scope}
	\end{tikzpicture}.
\end{equation*}
Resolving the rightmost strand and using the $\mathbb{Z}_{N}$ property (\ref{ZNproperty}) we may write
\begin{eqnarray*}
	e_{2i-1}^{(k)} e_{2i}^{(l)} e_{2i-1}^{(m)}&=&
	\omega^{l(k-m)} 
	\begin{tikzpicture}[anchorbase, scale =0.75]
		\boxcupcapbox{0}{~}{~}{$k$}{$-m$};
		\kboxw{2}{~}{$m-k$};
	\end{tikzpicture}
	=
	\omega^{l(k-m)} 
	\begin{tikzpicture}[anchorbase, scale =0.75]
		\boxcupcapbox{0}{~}{~}{$k-m$}{$-m$};
		\kbox{2}{~}{$m-k$};
		\begin{scope}[shift={(0,\genh)}]
			\kboxb{0}{~}{$m$};
			\sub{1}{~};
			\sub{2}{~};
		\end{scope}
	\end{tikzpicture}.
\end{eqnarray*}
Then applying the relation (\ref{cupboxes}) for the $m$-labelled cup gives
\begin{eqnarray*}
e_{2i-1}^{(k)} e_{2i}^{(l)} e_{2i-1}^{(m)}
	&=&
	\omega^{l(k-m)} 
	\omega^{(\frac{-N+k-m}{2})(k-m)}
	\begin{tikzpicture}[anchorbase, scale =0.75]
		\boxcupcapbox{0}{~}{~}{$m$}{$-m$};
		\su{2}{~};
		\begin{scope}[shift = {(0,\genh)}]
			\sub{0}{~};
			\kboxt{1}{~}{$k-m$};
			\kboxb{2}{~}{$m-k$};
		\end{scope}
	\end{tikzpicture}.
\end{eqnarray*}
By the isotopy relations (\ref{cupboxes})-(\ref{capboxes}) for a $k-m$ labelled box, we can identify the $\frac{\pi}{2}$ anti-clockwise rotation of the generator $e_{2i}^{(m-k)}$. 
And the diagram on the right can be associated with $\mathcal{F}^{-1}_{s} (e_{2i}^{(m-k)})$. It follows from (\ref{sftgens}) that 
\begin{equation}
	\begin{tikzpicture}[anchorbase, scale =0.75]
		\kboxtl{0}{~}{$k-m$};
		\kboxbl{1}{~}{~};
		\node at (2, \genh/3) {\scalebox{\scale}{$-(k-m)$}};
		\end{tikzpicture}
	= \frac{\omega^{-\big(\frac{N+k-m}{2}\big)(k-m)}}{\sqrt{N}}\sum_{p=1}^{N} \omega^{p(m-k) } 
	\begin{tikzpicture}[anchorbase, scale=0.75]
		\smallboxcupcapbox{1}{~}{~}{$-p$}{$p$};
	\end{tikzpicture}.
\end{equation}
The cubic relation then becomes
\begin{eqnarray}
e_{2i-1}^{(k)} e_{2i}^{(l)} e_{2i-1}^{(m)} 
&=&
  \frac{1}{\sqrt{N}}\sum_{p=1}^{N} \omega^{-(p-l)(k-m)}  
\begin{tikzpicture}[anchorbase, scale =0.75]
	\boxcupcapbox{0}{~}{~}{$m$}{$-m$};
	\su{2}{~};
	\begin{scope}[shift = {(0,\genh)}]
		\su{0}{~};
		\eboxcupcapbox{1}{~}{~}{$-p$}{$p$};
	\end{scope}
\end{tikzpicture}\\
&=&
\frac{1}{\sqrt{N}}\sum_{p=1}^{N}  \omega^{-(p-l)(k-m) }  e_{2i}^{(p)} e_{2i-1}^{(m)}.
\end{eqnarray}

A similar procedure is done for the $e_{2i}^{(k)} e_{2i-1}^{(l)} e_{2i}^{(m)}$ involving a clock wise $\pi/2$ rotation. 
Note by planar isotopy we may smoothly deform a horizontal strand into a cup or cap, and similarly, two horizontal lines into a cup/cap pair. This may then be treated as the $\pi/2$ rotation of the identity element operator. The string Fourier transform relation may then be applied to express the two horizontal strands as a sum over all labelled cups and caps. 

In general the clock model may be written in the presentation $\langle 1, e^{(0)}_{i}, \ldots, e^{(N-2)}_{i}|i \in \{1, \ldots, 2L-1 \}\rangle$ as 
%
%
\begin{eqnarray} 
	\label{clockhamdiagram} 
	H_{N} = &-\lambda \sum_{j=1}^{L} \sum_{k=0}^{N-2} ( \hat{\bar{\alpha}}_{k}  - \hat{\bar{\alpha}}_{-1} ) 	\begin{tikzpicture}[anchorbase, scale=0.6]
		\hboxcupcapbox{0}{$2i-1$}{$2i$}{$k$}{$- k$};
	\end{tikzpicture} - \sum_{j=1}^{L-1} \sum_{k=0}^{N-2} ( \hat{\alpha}_{k} - \hat{\alpha}_{-1})
	\begin{tikzpicture}[anchorbase, scale=0.6]
		\hboxcupcapbox{0}{$2i$}{$2i+1$}{$-k$}{$k$};
	\end{tikzpicture} \!\!.
\end{eqnarray}
Here the boundary term $- \sqrt{N} (L(\lambda +1)-1) \hat{\bar{\alpha}}_{1}$ has been omitted.

At $\lambda=1$ the Hamiltonian of the superintegrable chiral Potts chain may be written in the compact form 
\begin{equation}
	H_{\mathrm{SICP}} =\frac{1}{\sqrt{N}} \sum_{i=1}^{2L} \sum_{k=0}^{N-1}( \frac{1}{2}(N-1) + k )
	\begin{tikzpicture}[anchorbase, scale=0.75]
		\hboxcupcapbox{0}{$i$}{$i+1$}{$(-1)^{i+1} k$ }{$(-1)^{i} k$};
	\end{tikzpicture}.
\end{equation}
Here $H_{\mathrm{SICP}}= \frac{N}{4} (A_{0}+ \lambda A_{1})$ with the Dolan-Grady relations $\left[A_{0}, \left[A_{0} , \left[A_{0} , A_{1}\right]\right]\right] = 16 \left[A_{0}, A_{1}\right]$ and $\left[A_{1}, \left[A_{1},	\left[A_{1} , A_{0}\right]\right]\right] = 16 \left[A_{1}, A_{0}\right]$ satisfied by $A_{i} = \frac{4}{N} H_{i}$ for $i =0,1$ which may now be expressed diagrammatically as 
\begin{eqnarray}
	A_{1} &=& \frac{4}{N^{3/2}} \sum_{i=1}^{L} \sum_{k=0}^{N-1}( \frac{1}{2}(N-1) + k )
	\begin{tikzpicture}[anchorbase, scale=0.75]
		\hboxcupcapbox{0}{$2i-1$}{$2i$}{$k$}{$- k$};
	\end{tikzpicture},\\
    A_{0} &=& \frac{4}{N^{3/2}} \sum_{i=1}^{L} \sum_{k=0}^{N-1}( \frac{1}{2}(N-1) + k )
	\begin{tikzpicture}[anchorbase, scale=0.75]
		\hboxcupcapbox{0}{$2i$}{$2i+1$}{$-k$}{$k$};
	\end{tikzpicture}.
\end{eqnarray}

\subsection{Hilbert space description}
The planar parafermion algebra also allows us to give a diagrammatic description to the Hilbert space generalizing the notation of \cite{dGP} for the Temperley-Lieb algebra. Define a basis $\{ \ket{k}| k \in \mathbb{Z}_{N} \}$ with the `ket' states given by 
\begin{equation}
	\ket{k} = N^{- \frac{1}{4}}
	\raisebox{0.25cm}{
		\begin{tikzpicture}[anchorbase, scale=0.5]
			\boxcup{0}{}{}{$k$};
	\end{tikzpicture}}, \quad k = 0, \ldots, N-1,
\end{equation}
and for the adjoint `bra' states 
\begin{equation}
	\bra{k} =
	N^{- \frac{1}{4}}
	\raisebox{-0.125cm}{
		\begin{tikzpicture}[anchorbase, scale=0.5]
			\capbox{0}{}{}{$-k$};
	\end{tikzpicture}},  \quad k = 0, \ldots, N-1.
\end{equation}
In this basis we may write a representation of $\tau$ and $\sigma$ operators \cite{JaffeLiu} as  
\begin{equation}
	\tau =
	\begin{tikzpicture}[anchorbase, scale=0.75]
		\kboxhd{0}{}{$1$};
		\suhd{1}{};
	\end{tikzpicture}, \quad
    	\sigma = \omega^{\frac{N}{2}}
	\begin{tikzpicture}[anchorbase, scale=0.75]
		\kboxhd{0}{}{$1$};
		\kboxhd{1}{}{$-1$}
	\end{tikzpicture}.
\end{equation}
For $n = 0, \ldots, N-1$ we may also define a state with defects represented by straight strands as follows 
\begin{equation}
	\ket{k}_{i} =
	\raisebox{0.25 cm}{
		\begin{tikzpicture}[anchorbase, scale=0.5]
			\subt{0}{$1$};
			\node at (1, \genh/6) {$\cdots$};
			\subt{2}{};
			\boxcup{3}{$i$}{$i+1$}{$k$};
			\subt{5}{};
			\node at (6, \genh/6){$\cdots$};
			\subt{7}{$2L$};
	\end{tikzpicture}},
	\label{singlestateket}
\end{equation}
\begin{equation}
	\bra{k}_{i} =
	\raisebox{-0.125 cm}{
		\begin{tikzpicture}[anchorbase, scale=0.5]
			\sub{0}{$1$};
			\node at (1, \genh/6) {$\cdots$};
			\sub{2}{$1$};
			\capbox{3}{$i$}{$i+1$}{$-k$};
			\sub{5}{};
			\node at (6, \genh/6){$\cdots$};
			\sub{7}{$2L$};
	\end{tikzpicture}}.
	\label{singlestatebra}
\end{equation}
We may generalize such a basis to  $2L$ strands with $N^{L}$ states, $\{\ket{\vec{k}} \}$, with $\ket{\vec{k}} = \ket{k_{1}, \ldots, k_{L} }$ defined for $k_{i} \in \mathbb{Z}_{N}$ for $i = 1, \ldots, L$ as 
\begin{equation}
\label{ketstate}
	\ket{\vec{k}} =   \ket{k_{1}, \ldots, k_{L} } =  N^{- \frac{L}{4} }
	\raisebox{-0 cm}{
		\begin{tikzpicture}[anchorbase, scale=0.5]
			\boxcup{5}{}{}{$k_{L}$};
			\begin{scope}[shift={(0,\genh/3)}]	
				\boxcup{2}{}{}{$k_{2}$};
				\sub{5}{};
				\sub{6}{};
			\end{scope}
			\begin{scope}[shift={(0,2*\genh/3)}]	
				\boxcup{0}{}{}{$k_{1}$};
				\node at (4, 0) {$\cdots$};
				\sub{2}{};
				\sub{3}{};
				\sub{5}{};
				\sub{6}{};
			\end{scope}
	\end{tikzpicture}},
\end{equation}
and their adjoint states by 
\begin{equation}
	\bra{\vec{k}} =   \bra{k_{1}, \ldots, k_{L} } =  N^{- \frac{L}{4}}
	\raisebox{0 cm}{
		\begin{tikzpicture}[anchorbase, scale=0.5]
			\capbox{0}{}{}{$-k_{1}$};
			\sub{2}{};
			\sub{3}{};
			\sub{5}{};
			\sub{6}{};
			\begin{scope}[shift={(0,\genh/3)}]	
				\capbox{2}{}{}{$-k_{2}$};
				\node at (4, 0) {$\cdots$};
				\sub{5}{};
				\sub{6}{};
			\end{scope}
			\begin{scope}[shift={(0,2*\genh/3)}]	
				\capbox{5}{}{}{$-k_{L}$};	
			\end{scope}
	\end{tikzpicture}}. 
\end{equation}
Such a set of vectors forms an orthonormal basis of $N^{L}$ states with inner product given by \cite{JonesQuadraticTangles}
\begin{equation}
	\langle \vec{l}|\vec{k} \rangle =  N^{- \frac{L}{2}}
	\raisebox{ 0 cm}{
		\begin{tikzpicture}[anchorbase, scale=0.5]
			\capbox{0}{}{}{$-l_{1}$};
			\sub{2}{};
			\sub{3}{};
			\node at (4, 0) {$\cdots$};
			\sub{5}{};
			\sub{6}{};
			\begin{scope}[shift={(0,\genh/3)}]	
				\capbox{2}{}{}{$-l_{2}$};
				\sub{5}{};
				\sub{6}{};
			\end{scope}
			\begin{scope}[shift={(0,2*\genh/3)}]	
				\capbox{5}{}{}{$-l_{L}$};	
			\end{scope}
			\begin{scope}[shift={(0,-\genh/3)}]	
				\boxcup{0}{}{}{$k_{1}$};
				\sub{2}{};
				\sub{3}{};
				\sub{5}{};
				\sub{6}{};
			\end{scope}
			\begin{scope}[shift={(0,-2*\genh/3)}]	
				\boxcup{2}{}{}{$k_{2}$};
				\sub{5}{};
				\sub{6}{};
			\end{scope}
			\begin{scope}[shift={(0,-3*\genh/3)}]	
				\boxcup{5}{}{}{$k_{L}$};
			\end{scope}
	\end{tikzpicture}} = \delta_{k_{1}, l_{1}} \delta_{k_{2}, l_{2}} \cdots  \delta_{k_{L}, l_{L}}.
\end{equation}
This is an orthonormal set of vectors and spans the vector space \cite{JaffeLiu}.
Equivalently, by means of the twisted tensor product (\ref{tproduct}) we may write (\ref{ketstate}) as 
\begin{equation}
\label{pfbasis}
	\ket{\vec{k}} =  N^{- \frac{L}{4} } \omega^{\sum_{i=1}^{L-1} (\sum_{j=1}^{i} k_{j} ) k_{i-1}/2}
	\raisebox{0.25cm}{
		\begin{tikzpicture}[anchorbase, scale=0.5]
			\boxcup{0}{}{}{$k_{1}$};
			\boxcup{2}{}{}{$k_{2}$};
			\node at (4, 0.5) {$\cdots$};
			\boxcup{6}{}{}{$k_{L}$};
	\end{tikzpicture}}.
\end{equation}

\section{Staggered XXZ and the coupled TL algebra}

In this section we present an algebraic and pictorial presentation of an algebra related to the coupled TL algebra. It was shown in \cite{FM} that one may define a representation related to the XXZ spin chain Hamiltonian at $q=1$ satisfying a similar algebra to the coupled TL algebra (\ref{ctlrel1})-(\ref{ctlrel4}).
The Hamiltonian on $L$ sites admits a representation of $\mathrm{TL}_{L}(q)$ in the form of (\ref{tlham}) with $\mathrm{SU}_{q}(2)$ boundary conditions \cite{Nichols2006}
\begin{equation}
	\label{xxztlgen}
	e_{i} =  - \frac{1}{2} \left( \sigma_{i}^{x}\sigma_{i+1}^{x} + \sigma_{i}^{y}\sigma_{i+1}^{y}  + \cos \gamma \sigma_{i}^{z}\sigma_{i+1}^{z} - \cos \gamma+ \mathrm{i} \sin \gamma (\sigma_{i}^{z} - \sigma_{i+1}^{z}) \right).
\end{equation}
%
%
Here $q = e^{\mathrm{i} \gamma}$ and $\sigma_{j}^{\alpha}$ for $\alpha =x,y,z$ are the Pauli spin matrices acting on site $j$ of $(\mathbb{C}^{2})^{\otimes L}$ \footnote{Here $q$ is unrelated to the $q$ of \cite{JaffeLiu}.}.
The anisotropy parameter of the Hamiltonian is given by $\Delta=-\cos \gamma$.
For $i = 1, \ldots, L-1$ define
\begin{eqnarray}\label{sxxrep}
	e^{(0)}_{i} &=& \frac{1}{2}(1- \sigma_{i}^{z} \sigma_{i+1}^{z}  +  \sigma_{i}^{x} \sigma_{i+1}^{x} + \sigma_{i}^{y} \sigma_{i+1}^{y} ), \label{sxx1} \\
	e^{(1)}_{i} &=& \frac{1}{2}(1- \sigma_{i}^{z} \sigma_{i+1}^{z}  -  \sigma_{i}^{x} \sigma_{i+1}^{x} - \sigma_{i}^{y} \sigma_{i+1}^{y} ), \label{sxx2} 
\end{eqnarray}
satisfying for $k,l=0,1$
\begin{eqnarray}
\label{sxxrelation1}
e_{i}^{(k)} e_{i}^{(l)} &=& 2 \delta_{k,l} e_{i}^{(k)}, \quad 
e_{i}^{(k)} e_{i \pm 1}^{(l)} e^{(k)}_{i} = e_{i}^{(k)},\\
e_{i}^{(k)} e_{j}^{(l)} &=& e_{j}^{(l)} e_{i}^{(k)}, \quad|i-j| \ge 2.
\end{eqnarray}
The relations above reduce to the TL relations for a single value of $k$. Here $e_{i}^{(1)}$ is equivalent to the XXZ TL representation (\ref{xxztlgen}) at $q=1$. The operators (\ref{sxx1})-(\ref{sxx2}) satisfy additional cubic relations similar to (\ref{cubicrel1})-(\ref{cubicrel2}), 
\begin{eqnarray}
\label{sxxcubics1}
	e^{(1)}_{i} e^{(1)}_{i \pm 1} e_{i}^{(0)} &=& e^{(1)}_{i} e^{(0)}_{i \pm 1} e^{(0)}_{i},\quad
	e_{i}^{(0)} e^{(1)}_{i \pm 1} e^{(1)}_{i} = e^{(0)}_{i} e^{(0)}_{i \pm 1} e^{(1)}_{i}.
\end{eqnarray}
Along with the relations 
\begin{eqnarray}
	e^{(1)}_{i} e^{(0)}_{i \pm 1} e^{(0)}_{i} &=&  e^{(0)}_{i \pm 1} e^{(0)}_{i} + e^{(1)}_{i \pm 1} e^{(0)}_{i}  -  e^{(0)}_{i},\label{sxxcubicrel1}\\
	&=&  e^{(1)}_{i} e^{(0)}_{i \pm 1}  + e^{(1)}_{i} e^{(1)}_{i \pm 1} -  e^{(1)}_{i}, \label{sxxcubicrel2} \\
	e^{(0)}_{i} e^{(0)}_{i \pm 1} e^{(1)}_{i} &=&  e^{(0)}_{i \pm 1} e^{(1)}_{i} + e^{(1)}_{i \pm 1} e^{(1)}_{i} - e^{(1)}_{i},\label{sxxcubicrel3}\\
	&=& e^{(0)}_{i} e^{(0)}_{i \pm 1}  + e^{(0)}_{i} e^{(1)}_{i \pm 1}  - e^{(0)}_{i}.\label{sxxcubicrel4}
\end{eqnarray}

The Hamiltonian of the staggered XX chain is given in terms of two parameters $\lambda_{1}, \lambda_{2}$ by
\begin{eqnarray}
\label{sxxhamiltonian}
	H_{\mathrm{sXX}} &=& \sum_{i} \lambda_{1} (e^{(0)}_{2i} - e^{(1)}_{2i}) + \lambda_{2} (e^{(0)}_{2i+1} - e^{(1)}_{2i +1}).
\end{eqnarray}
It is also pointed out in \cite{FM} that one may write the Hamiltonian (\ref{sxxhamiltonian}) as $H = \lambda_{1} A_{0} + \lambda_{2} A_{1} $, with $A_{0}$ and $A_{1}$ the generators of an Onsager algebra, the additional integrable structure of the superintegrable chiral Potts model. In terms of the generators (\ref{sxx1})-(\ref{sxx2}), $A_{0}$ and $A_{1}$ are given by
\begin{equation}
	\label{sxxonsager}
	A_{0} =\sum_{i} (e^{(0)}_{2i} - e^{(1)}_{2i}), \quad A_{1} = \sum_{i} (e^{(0)}_{2i+1} - e^{(1)}_{2i+1}),
\end{equation}
satisfying the Dolan-Grady conditions for an even number of sites $L$ and under periodic boundary conditions.

\subsection{Pictorial Representation}
Similar to the planar para algebra construction for the $\mathrm{cTL}_{n}(\sqrt{N})$ representation related to the $\mathbb{Z}_{N}$-clock model we may construct an analogous diagrammatic representation for the operators (\ref{sxx1})-(\ref{sxx2}).
In this representation we still expect the $\pi/2$-rotation of the cup and cap to be the identity object and to play a role in the cubic relations (\ref{sxxcubicrel1})-(\ref{sxxcubicrel4}).
We can relate the two generators (\ref{sxx1}) and (\ref{sxx2}) by a conjugation by the Pauli spin matrix $\sigma^{z}$, $e^{(1)}_{i} = \sigma^{z} e^{(0)}_{i} \sigma^{z}$.
Here we note the relations $\sigma^{z}_{i} e_{i}^{(k)} = - \sigma^{z}_{i+1} e_{i}^{(k)}$ and $e_{i}^{(k)} \sigma^{z}_{i} = - e_{i}^{(k)}\sigma^{z}_{i+1} $, equivalent to moving a 1-box across a cup or cap respectively. 
Denoting the operation of $\sigma^{z}_{i}$ on the $i$-th site of the Hilbert space $(\mathbb{C}^{2} )^{\otimes L}$ as $\mathbb{Z}_{2}$-graded 1-box acting on the $i$-th strand with multiplication treated modulo 2 
\begin{equation}
	\sigma_{i}^{z}  := 	
	\raisebox{-0.25cm}{
		\begin{tikzpicture}[anchorbase, scale=0.75]
			\kboxhd{1}{$i$}{$1$};
	\end{tikzpicture}}, \quad 
	\raisebox{-0.25cm}{
	\begin{tikzpicture}[anchorbase, scale=0.75]
		\kboxh{0}{$i$}{$1$};
		\begin{scope}[shift={(0, \genh/3)}]
			\kboxh{0}{}{$1$};
		\end{scope}
\end{tikzpicture}}
= 
\raisebox{-0.25cm}{
	\begin{tikzpicture}[anchorbase, scale=0.75]
		\suhd{0}{$i$};
\end{tikzpicture}}, \quad 
	\raisebox{-0.25cm}{
	\begin{tikzpicture}[anchorbase, scale=0.75]
		\sub{0}{$i$};
		\kboxh{1}{$j$}{$1$};
		\begin{scope}[shift={(0, \genh/3)}]
			\sub{1}{};
			\kboxh{0}{}{$1$};
		\end{scope}
\end{tikzpicture}}
= 
	\raisebox{-0.25cm}{
	\begin{tikzpicture}[anchorbase, scale=0.75]
		\kboxh{0}{$i$}{$1$};
		\sub{1}{$j$};
		\begin{scope}[shift={(0, \genh/3)}]
			\sub{0}{};
			\kboxh{1}{}{$1$};
		\end{scope}
\end{tikzpicture}}.
\end{equation}
Unlike the parafermionic 1-boxes these commute on different sites 
and satisfy the following  the isotopy relations, where contractable loops take value $\delta = 2$,
\begin{equation}
    \begin{tikzpicture}[anchorbase,scale=0.75]
		\draw[usual] (0,0) circle (\genh/6);
		\filldraw[usual, fill = boxfill] (-\genh/6 - \genh/24, -\genh/24) rectangle (-\genh/6 + \genh/24, \genh/24);
		\node[anchor=east] at (-\genh/6, 0){\scalebox{\scale}{$k$}};
	\end{tikzpicture}
	= 2\delta_{k,0}, \qquad 
    (-1)^{k}
	\begin{tikzpicture}[scale=.5, anchorbase]
		\bboxcup{0}{~}{~}{$k$};
		\sub{2}{};
		\begin{scope}[shift={(0, \genh/3)}]
			\shortcap{1}{~}{~};
			\sub{0}{~};
		\end{scope}
	\end{tikzpicture}
	= 
		\begin{tikzpicture}[scale=.5, anchorbase]
		\kboxhd{0}{}{$ k $};
	\end{tikzpicture}
		= (-1)^{k}
	\begin{tikzpicture}[scale=.5, anchorbase]
		\sub{-1}{};
		\boxcup{0}{~}{~}{$k$};
		\begin{scope}[shift={(0, \genh/3)}]
			\shortcap{-1}{~}{~};
			\sub{1}{};
		\end{scope}
	\end{tikzpicture},
\end{equation}
with $k=0,1$. 
We may now make the identification for the generators of $\mathrm{cTL}_{n}(\sqrt{N})$. Denoting $e^{(0)}_{i}$ by the usual cup and cap diagram and $e^{(1)}_{i}$ by a generic 2-box acting on the $i$th and $i+1$th strands. We may denote the 2-box as the conjugation of $\sigma^{z}$ 1-boxes as follows
\begin{equation}
	e^{(0)}_{i} = 
	\raisebox{-0.25cm}{
		\begin{tikzpicture}[anchorbase, scale=0.75]
			\cupcaphb{0}{$i$}{$i+1$};
	\end{tikzpicture}}, \quad 
    	e^{(1)}_{i} = 
	\raisebox{-0.25cm}{	
		\begin{tikzpicture}[anchorbase, scale=0.75]
			\suhb{0}{};
			\suhb{1}{};
			\twoboxs{0}{$1$}{};
			\node at (0,-.5) {\scalebox{\scale}{$i$}};
			\node at (0+1,-.5) {\scalebox{\scale}{$i+1$}};
		\end{tikzpicture}
	}
	:=
	\raisebox{-0.25cm}{
		\begin{tikzpicture}[anchorbase, scale=0.75]
			\boxcupcapboxhb{0}{$i$}{$i+1$}{$1$}{$1$};
	\end{tikzpicture}},
\end{equation}
to obtain a pictorial representation of (\ref{sxx1})-(\ref{sxx2}). 
We note the construction is similar to the framization of the TL algebra \cite{FrameTL}. 
The one-strand rotation in this pictorial representation does not correspond to an algebraic Fourier transform over $\mathbb{Z}_{N}$ as in (\ref{sftgens}), instead we note that $e^{(0)}_{i} + e^{(1)}_{i} = (1 - \sigma_{i}^{z} \sigma_{i+1}^{z})$, i.e., 
\begin{equation}
	\label{sxxskeindiagram}
	\raisebox{-0.25cm}{
		\begin{tikzpicture}[anchorbase, scale=0.7]
			\cupcaphb{0}{}{};
	\end{tikzpicture}}
	+ 
	\raisebox{-0.25cm}{
		\begin{tikzpicture}[anchorbase, scale=0.7]
			\boxcupcapboxhb{0}{}{}{$1$}{$1$};
	\end{tikzpicture}}
	= 
	\raisebox{-0.25cm}{
		\begin{tikzpicture}[anchorbase, scale=0.75]
			\sub{0}{};
			\sub{1}{};
			\begin{scope}[shift={(0, \genh/3)}]
				\sub{0}{};
				\sub{1}{};
			\end{scope}
	\end{tikzpicture}}
	- 
	\raisebox{-0.25cm}{
		\begin{tikzpicture}[anchorbase, scale=0.75]
			\sub{0}{};
			\kboxh{1}{}{$1$};
			\begin{scope}[shift={(0, \genh/3)}]
				\sub{1}{};
				\kboxh{0}{}{$1$};
			\end{scope}
	\end{tikzpicture}}.
\end{equation}
That is, $\mathcal{F}_{s} (e^{(1)}_{i} ) = 1 - (e^{(0)}_{i} + e^{(1)}_{i})$.
Next, we write the rotated 2-box $e^{(1)}_{i}$ by the relation (\ref{sxxskeindiagram})
\begin{equation}
	-	\raisebox{-0.25cm}{
		\begin{tikzpicture}[anchorbase, scale=0.75]
			\sub{0}{};
			\kboxh{1}{}{$1$};
			\begin{scope}[shift={(0, \genh/3)}]
				\sub{1}{};
				\kboxh{0}{}{$1$};
			\end{scope}
	\end{tikzpicture}}
= 
	\raisebox{-0.25cm}{
		\begin{tikzpicture}[anchorbase, scale=0.7]
			\twoboxrs{1}{$1$}{};
			\node at (1,-.5) {\scalebox{\scale}{}};
			\node at (1+1,-.5) {\scalebox{\scale}{}};
	\end{tikzpicture}}
	=
	\raisebox{-0.25cm}{
		\begin{tikzpicture}[anchorbase, scale=0.7]
			\cupcaphb{0}{}{};
	\end{tikzpicture}}
	+ 
	\raisebox{-0.25cm}{
		\begin{tikzpicture}[anchorbase, scale=0.7]
			\boxcupcapboxhb{0}{}{}{$1$}{$1$};
	\end{tikzpicture}}
	-
	\raisebox{-0.25cm}{
		\begin{tikzpicture}[anchorbase, scale=0.75]
			\sub{0}{};
			\sub{1}{};
			\begin{scope}[shift={(0, \genh/3)}]
				\sub{0}{};
				\sub{1}{};
			\end{scope}
	\end{tikzpicture}}.
\end{equation}
The cubic relations in the sXX case follow from 
\begin{equation}
	e^{(1)}_{i} e^{(0)}_{i+1} e^{(0)}_{i} = 
	\begin{tikzpicture}[anchorbase, scale=0.75]
		\cupcapb{0}{}{};
		\sub{2}{};
		\begin{scope}[shift={(0, \genh/3)}]
			\sub{0}{};
			\cupcapb{1}{}{};
		\end{scope}
		\begin{scope}[shift={(0, 2*\genh/3)}]
			\sub{0}{};
			\sub{1}{};
			\sub{2}{};
			\twoboxb{0}{$1$}{};
		\end{scope}
		\begin{scope}[shift={(0, \genh)}]
			\sub{0}{};
			\sub{1}{};
			\sub{2}{};
		\end{scope}
	\end{tikzpicture}
	=
	\begin{tikzpicture}[anchorbase, scale=0.75]
		\cupcapb{0}{}{};
		\sub{2}{};
		\begin{scope}[shift={(0, \genh/3)}]
			\su{0}{};
			\twoboxr{1}{$1$}{};
		\end{scope}
	\end{tikzpicture},
	\label{sxxcubicproof}
\end{equation}
substituting into the relation (\ref{sxxcubicproof}) yields the cubic relation 
\begin{equation}
	e^{(1)}_{i} e^{(0)}_{i+1} e^{(0)}_{i} = 
	\begin{tikzpicture}[anchorbase, scale=0.75]
		\cupcapb{0}{}{};
		\sub{2}{};
		\begin{scope}[shift={(0, \genh/3)}]
			\suhb{0}{};
			\cupcaphb{1}{}{};
		\end{scope}
	\end{tikzpicture}
	+
	\begin{tikzpicture}[anchorbase, scale=0.75]
		\cupcapb{0}{}{};
		\sub{2}{};
		\begin{scope}[shift={(0, \genh/3)}]
			\suhb{0}{};
			\boxcupcapboxhb{1}{}{}{$1$}{$1$};
		\end{scope}
	\end{tikzpicture}
	-
	\begin{tikzpicture}[anchorbase, scale=0.75]
		\cupcapb{0}{}{};
		\sub{2}{};
		\begin{scope}[shift={(0, \genh/3)}]
			\suhb{0}{};
			\suhb{1}{};
			\suhb{2}{};
		\end{scope}
	\end{tikzpicture}.
	\label{sxxcubicproof2}
\end{equation}
Here the right hand side may be identified as $ e^{(0)}_{i+1} e^{(0)}_{i} + e^{(1)}_{i+1}e^{(0)}_{i} - e^{(0)}_{i}$, with a similar calculation for the $e^{(1)}_{i} e^{(0)}_{i-1} e^{(0)}_{i}$ case. 

The XXZ Hamiltonian on $L$ sites under periodic boundary conditions is given by 
\begin{equation}
	\label{hsp}
	H_{\mathrm{XXZ}}= \frac{1}{2} \sum_{i}^{L} \left( \sigma_{i}^{x} \sigma_{i+1}^{x}  + \sigma_{i}^{y} \sigma_{i+1}^{y} + \Delta \left( 1+ \sigma_{i}^{z} \sigma_{i+1}^{z} \right) \right).
\end{equation}
The Hamiltonian of the XXZ spin chain may be recast in terms of the generators of the coupled TL algebra 
\begin{equation}
	H = \sum_{i}^{L} ( \frac{1}{2}(e^{(0)}_{i} - e^{(1)}_{i})+ \Delta (1- \frac{1}{2} (e^{(0)}_{i} + e^{(1)}_{i}) ) ).
\end{equation}
Here the parameter $\Delta$ is now independent of the coupled TL algebra.
The components of the Hamiltonian may be written as 
\begin{equation}
	S_{i} = \frac{1}{2} (e^{(0)}_{i} - e^{(1)}_{i}) , \quad P_{i} = 1 - \frac{1}{2} (e^{(0)}_{i} + e^{(1)}_{i}).
\end{equation}
The operators $S_{i}, P_{i}$ form a representation of  a chromatic algebra $\bra{S_{i}, P_{i}} \ket{i \in \{1, \ldots, n\}}$ introduced in \cite{EckRydberg} defined by the relations
\begin{eqnarray}
\label{chromaticalgebra}
	&&S_{i}^{2} = 1 - P_{i}, \quad P_{i}^{2} = P_{i}, \quad S_{i} P_{i} = P_{i} S_{i} = 0,\\
    &&\qquad \qquad S_{i}S_{i \pm 1} S_{i} = P_{i} S_{i \pm 1} P_{i} = 0,  \nonumber
\end{eqnarray}
with far-apart commutation between generators
\begin{equation}
	S_{i} S_{j} = S_{j} S_{i}, \quad S_{i} P_{j} = P_{j} S_{i}, \quad P_{i} P_{j} = P_{j} P_{i}, \quad |i-j| \ge 2. 
\end{equation}
The representation (\ref{sxx1})-(\ref{sxx2}) gives 
\begin{eqnarray}
	e^{(0)}_{i} + e^{(1)}_{i} &=& 1 - \sigma^{z}_{i} \sigma^{z}_{i+1}, \\
	e^{(0)}_{i} - e^{(1)}_{i} &=& \sigma^x_i \sigma^x_{i+1} +\sigma^y_i \sigma^y_{i+1},
\end{eqnarray}
and $S_{i}, P_{i}$ are given by
\begin{equation}
	\label{chromaticxxz}
	S_{i} = \frac{1}{2} \left( \sigma_{i}^{x} \sigma_{i+1}^{x}  + \sigma_{i}^{y} \sigma_{i+1}^{y} \right), \quad P_{i} = \frac{1}{2} \left( 1+ \sigma_{i}^{z} \sigma_{i+1}^{z} \right).
\end{equation}

The chromatic algebra (\ref{chromaticalgebra}) admits a diagrammatic presentation in terms of trivalent planar graphs
\begin{equation}
\label{chromaticpic}
	P_{i} = 
	\raisebox{-0.25cm}{
		\begin{tikzpicture}[anchorbase, scale=0.75]
			\pgen{0}{}{};
	\end{tikzpicture}},\quad 
	S_{i} = 
	\raisebox{-0.25cm}{
		\begin{tikzpicture}[anchorbase, scale=0.75]
			\sgen{0}{}{};
	\end{tikzpicture}}. 
\end{equation}
The generators are related to a TL generator $E_{i}$ via the contraction-deletion property $P_{i} + E_{i} = S_{i} + 1$
\begin{equation}
	\raisebox{-0.25cm}{
		\begin{tikzpicture}[anchorbase, scale=0.75]
			\pgen{0}{}{}
	\end{tikzpicture}}
	+
	\raisebox{-0.25cm}{
		\begin{tikzpicture}[anchorbase, scale=0.75]
			\egen{0}{}{};
	\end{tikzpicture}}
	=
	\raisebox{-0.25cm}{
		\begin{tikzpicture}[anchorbase, scale=0.75]
			\sgen{0}{}{}
	\end{tikzpicture}}
	+
	\raisebox{-0.25cm}{
		\begin{tikzpicture}[anchorbase, scale=0.75]
			\idgen{0}{}{};
	\end{tikzpicture}}.
\end{equation}
A closed loop contributes a factor of $Q-1$ and a closed loop with a single strand attached vanishes
\begin{equation}
	\raisebox{-0.0cm}{
		\begin{tikzpicture}[anchorbase, scale=0.75]
			\draw[usual] (0,0) circle (0.5);
	\end{tikzpicture}}
	= (Q-1), \quad 
	\raisebox{-0.0cm}{
		\begin{tikzpicture}[anchorbase, scale=0.75]
			\draw[usual] (-1,0) to (-0.5,0);
			\draw[usual] (0,0) circle (0.5);
	\end{tikzpicture}}
	= 0.
\end{equation}
At $Q=3$ the generators (\ref{chromaticpic}) satisfy the relations (\ref{chromaticalgebra}). 
The repeated application of which reduces a trivalent graph, with no free strands, to a sum over closed loops and loops with one strand. 
The result determines the polynomial $\chi_{\mathcal{G}}(Q)/Q$, where $\chi_{\mathcal{G}}(Q)$ is the chromatic polynomial giving the number of ways a planar graph $\mathcal{G}$ with $Q$ colours may be coloured, with the restriction that two neighboring regions must differ in colour. 
We may write $E_{i} = e_{i}^{(0)}$ with loop value $Q-1=2$ and the generators of the chromatic algebra in terms of those of $\mathrm{cTL}_{n}(\sqrt{N})$. For the $P_{i}$ we may write 
\begin{equation}
	\raisebox{-0.25cm}{
		\begin{tikzpicture}[anchorbase, scale=0.75]
			\pgen{0}{}{};
	\end{tikzpicture}} =
	\raisebox{-0.25cm}{
		\begin{tikzpicture}[anchorbase, scale=0.75]
			\sub{0}{};
			\sub{1}{};
			\begin{scope}[shift={(0, \genh/3)}]
				\sub{0}{};
				\sub{1}{};
			\end{scope}
	\end{tikzpicture}}
	- \frac{1}{2} 
		\raisebox{-0.25cm}{
		\begin{tikzpicture}[anchorbase, scale=0.7]
			\cupcaphb{0}{}{};
	\end{tikzpicture}}
	- \frac{1}{2} 
	\raisebox{-0.25cm}{
		\begin{tikzpicture}[anchorbase, scale=0.7]
			\boxcupcapboxhb{0}{}{}{$1$}{$1$};
	\end{tikzpicture}},
\end{equation}
and for the $S_{i}$ generator 
\begin{equation}
	\raisebox{-0.25cm}{
		\begin{tikzpicture}[anchorbase, scale=0.75]
			\sgen{0}{}{};
	\end{tikzpicture}}
= \frac{1}{2} 
	\raisebox{-0.25cm}{
	\begin{tikzpicture}[anchorbase, scale=0.7]
		\cupcaphb{0}{}{};
\end{tikzpicture}}
-\frac{1}{2} 
\raisebox{-0.25cm}{
	\begin{tikzpicture}[anchorbase, scale=0.7]
		\boxcupcapboxhb{0}{}{}{$1$}{$1$};
\end{tikzpicture}}.
\end{equation}
From which one can derive additional relations of the algebra such as $P_{i}S_{i} = 0$.
We note that by the relation (\ref{sxxskeindiagram}) the $\pi/2$ rotation of $P_{i}$ gives $S_{i}$. 

\section{Discussion}

In this article we have provided the $N$-state generalization of the coupled TL algebra presented in the $N=3$ case by Fjelstad and M\r{a}nsson~\cite{FM}. The diagrammatic description of this algebra involves parafermionic operators attached to strands of a planar algebra where rotations may be resolved via a Fourier transform relation. 
The string Fourier transform (\ref{sft}) has the action given by (\ref{sftgens}), which leads to the correct cubic relations (\ref{ctlrel3})-(\ref{ctlrel4}).
The planar algebra $\mathrm{PF}_{n}$ provides the correct framework for a description of both the Hamiltonian and the Hilbert space. 
A generalization of the rotation action of the string Fourier transform also provides the correct pictorial description of the staggered XX representation, and has also been shown to describe a chromatic algebra related to a link invariant of trivalent graphs. 

It remains an interesting open question as to if the coupled TL algebra plays a role as a spectrum generating algebra of a corresponding Hamiltonian. 
For example, in the case of the usual TL algebra, along with the pictorial representation, one may derive the full eigenspectrum of the TL Hamiltonian, in that case via the Bethe Ansatz, as done, e.g., in Refs~\cite{L90, L91, MS, dGP, AK, NP}. 
The question then is if the SICP eigenspectrum can be obtained via the coupled TL algebra and pictorial representation given here. 
Notably, for periodic boundary conditions, although the structure of the spectrum is determined from the Onsager algebra, the Baxter polynomials inherent to its solution are not obtainable via the Hamiltonian alone. 
These polynomials are related to a type of generalized Chebyshev polynomials.
It is an interesting open question as to if the algebra presented here plays a role in this direction, particularly as a generalization of the affine Temperley-Lieb algebra \cite{GL1998, QW}. 
Similarly one can also consider the version for open boundary conditions, where there is no known solution for the SICP Hamiltonian. 
A related issue is finding other possible representations of the coupled algebra.
As observed by Fjelstad and M\r{a}nsson the components of the staggered XX Hamiltonian satisfy the Dolan-Grady relations and hence generate an Onsager algebra. In our calculations it appears the relations of the coupled TL algebra reduce the Dolan-Grady relation to an equivalent relation involving the generators of the coupled algebra. One may hope to find other integrable models possessing an Onsager structure via a representation of the coupled TL algebra.

The description of the states Hilbert space in $\mathrm{PF}_{n}$, appears to generalize that of the usual TL algebra, where the Hilbert space decomposes into sectors $\mathcal{W}_{j}$ with $2j$ free strands or `defects' \cite{Westbury, RSA}.
This is generalized in the `blob' algebra~\cite{MS-blob,dGN} where such defects are also allowed to carry additional idempotent operators of the algebra. 
We expect a similar decomposition for the coupled TL algebra where the strands in the link states of the Hilbert space carry additional parafermionic operators.
An additional open question is to the dimension of the coupled TL algebra for both even and odd $n$. It appears that with the identity relation in the even strand case, $n=2L$, the dimension of $\mathrm{cTL}_{n}(\sqrt{N})$ is $N^{2L-1}$.

We also note that the planar parafermion algebra possess a generalization of the Gaussian representation of the braid group \cite{JaffeLiu}. 
Such a braid appears to define a representation of the BMW algebra up to $N=5$, satisfying additional parafermion commutation relations. 
Such relations may be used to express all generators of the coupled TL algebra in terms of crossings and a single $\mathbb{Z}_{N}$ graded 1-box on the first strand, generalizing the original pictorial representation of $\mathrm{cTL}_{n}(\sqrt{N})$ in the $N=3$ case including a pole \cite{ABW}. 
Such observations are to be followed up in a later article \cite{AB2025}.

~

\noindent
This paper is dedicated to the memory of our colleague and mentor, Rodney James Baxter.

\ack 
The authors thank Hoel Queffelec for many insightful discussions and Xavier Poncini and J{\o}rgen Rasmussen for sharing and discussing their unpublished work. 
We thank the mathematical research institute MATRIX in Australia where part of this research was performed.  
This work has been supported by Australian Research Council Grants DP210102243 and DP240100838.

\end{document}